\documentclass[
 twocolumn,
superscriptaddress,
 amsmath,amssymb,
 aps,
 pra,
 floatfix,
]{revtex4-2}
\usepackage[normalem]{ulem}
\usepackage{graphicx}
\usepackage{dcolumn}
\usepackage{bm}
\usepackage[colorlinks]{hyperref}
\hypersetup{%
	plainpages=true,
	breaklinks=true,
	hypertexnames=false,
	pageanchor=true,
	colorlinks=true,
	linkcolor={blue},
	citecolor={red},
	urlcolor={blue},
	anchorcolor={black}
}

\usepackage{physics}
\usepackage{xcolor}
\usepackage{soul}

\newcommand{\sz}{\hat \sigma_z}
\newcommand{\sy}{\hat \sigma_y}
\newcommand{\sx}{\hat \sigma_x}
\newcommand{\sm}{\hat \sigma_-}
\renewcommand{\sp}{\hat \sigma_+}

\newcommand{\aop}{\hat a}
\newcommand{\adop}{\hat a ^\dagger}

\newcommand{\figref}[1]{\mbox{Fig.~\ref{#1}}}

\newcommand{\secref}[1]{\mbox{Sec.~\ref{#1}}}

\newcommand{\appref}[1]{\mbox{Appendix~\ref{#1}}}
\renewcommand{\eqref}[1]{\mbox{Eq.~(\ref{#1})}}

\newcommand{\be}{\begin{equation}}
\newcommand{\ee}{\end{equation}}
\newcommand{\bea}{\begin{eqnarray}}
\newcommand{\eea}{\end{eqnarray}}

\usepackage{xcolor}

\definecolor{darkgreen}{rgb}{0.1, 0.3, 0.0}

\newcommand{\figpanel}[2]{Fig.~\hyperref[#1]{\ref*{#1}(#2)}}
\newcommand{\figpanels}[3]{Fig.~\hyperref[#1]{\ref*{#1}(#2)-(#3)}}
\newcommand{\figpanelNoPrefix}[2]{\hyperref[#1]{\ref*{#1}(#2)}}

\begin{document}

\title{ Regimes of Cavity-QED under Incoherent Excitation: From Weak to Deep Strong Coupling}
\author{Alberto Mercurio}
\email{alberto.mercurio@unime.it }
\affiliation{Dipartimento di Scienze Matematiche e Informatiche, Scienze Fisiche e  Scienze della Terra, Universit\`{a} di Messina, I-98166 Messina, Italy}
\author{Vincenzo Macr\`{i}}
\email{vincenzo.macri@riken.jp}
\affiliation{Theoretical Quantum Physics Laboratory, RIKEN, Wako-shi, Saitama 351-0198, Japan}


\author{Chris Gustin}
\affiliation{Department of Physics, Engineering Physics and Astronomy, Queen's University, Kingston, ON K7L 3N6, Canada}
\affiliation{Department of Applied Physics, Stanford University, Stanford, California 94305, USA}


\author{Stephen Hughes}
\affiliation{Department of Physics, Engineering Physics and Astronomy, Queen's University, Kingston, ON K7L 3N6, Canada}

\author{Salvatore Savasta}
\affiliation{Theoretical Quantum Physics Laboratory, RIKEN, Wako-shi, Saitama 351-0198, Japan}
\affiliation{Dipartimento di Scienze Matematiche e Informatiche, Scienze Fisiche e  Scienze della Terra, Universit\`{a} di Messina, I-98166 Messina, Italy}

\author{Franco Nori}
\affiliation{Theoretical Quantum Physics Laboratory, RIKEN, Wako-shi, Saitama 351-0198, Japan}
\affiliation{Physics Department, The University of Michigan, Ann Arbor, Michigan 48109-1040, USA}

\date{\today}

\begin{abstract}
The prototypical system constituted by 
a two-level atom interacting with a quantized single-mode electromagnetic field is described by the quantum Rabi model (QRM).  The QRM  is potentially valid at any light-matter interaction regime, ranging from the weak (where the decay rates exceeds the coupling rate)  to the deep strong coupling (where the interaction rate exceeds the bare transition frequencies of the subsystems). However, when  reaching the ultrastrong coupling regime, several theoretical issues may prevent the correct description of the observable dynamics of such a system: (i) the standard quantum optics master equation fails to correctly describe the interaction of this system with the reservoirs;  (ii) the correct output photon rate is no longer proportional to the intracavity photon  number; and (iii) the appears to violate gauge invariance. Here, we study the photon flux emission rate  
of this system under the incoherent excitation of the two-level atom for any light-matter interaction strength, and consider different effective temperatures. 
The dependence of the emission spectra  on the coupling strength is the result of the interplay between energy levels, matrix elements of the observables, and the density of states of the reservoirs.
Within this approach, we also study the occurence of light-matter decoupling in the deep strong coupling regime, and show how
all of the 
obtained results are gauge invariant.
 
\end{abstract}

\maketitle

\section{Introduction}
\label{sec: Introduction}


The quantum Rabi model (QRM) provides the simplest full quantum description of light-matter interaction. It is also one of the most well studied models in quantum optics, and a cornerstone of cavity quantum electrodynamics (cavity QED). The quantum Rabi Hamiltonian describes the dipolar interaction of a two-level system (TLS) or qubit with a single quantized mode of an electromagnetic resonator. It was first introduced by Jaynes and Cummings \cite{Jaynes1963}, in order to compare the semiclassical model, previously introduced by Rabi \cite{Rabi1936}, with a full quantum model. To solve the model, the rotating wave approximation (RWA) is commonly introduced. In this approximation, known as the Jaynes-Cummings model (JCM), the counter-rotating terms are neglected. This is a valid approximation for the near resonance case and when the light-matter coupling rate is much smaller than the resonance frequency of the TLS, or equivalently of the quantized cavity mode. These conditions are fulfilled in several experimental settings \cite{Kockum2019Rev,Forn-Diaz2019}. The RWA may also fail for describing the optical pumping,
when the excitation field is sufficiently strong
or for a nonlinear short pulse excitations with several carrier cycles~\cite{PhysRevLett.81.3363}.

Despite its simplicity, the QRM is able to describe a wide variety of light-matter quantum systems and gives rise to a great diversity of behaviors and effects, depending on the relative magnitude of the light-matter coupling strength.
However, the description and the analysis of experiments on systems which can be potentially modeled by the quantum Rabi Hamiltonian require additional theoretical tools. For example, it is necessary to take into account the interaction of both the TLS and cavity photons with the external environment (thermal reservoirs) \cite{Haroche2006}. Of course, in the absence of such interactions, basic features such as the excitation of the system components, dissipation and decoherence effects, and the detection of photons outside the cavity cannot be described properly. Even the definition of the different light-matter interaction regimes requires one to include the interaction of the system components with their reservoirs. 

In  weak coupling regime of cavity QED,
dissipation is stronger than the intrinsic coherent coupling between matter and the light.
Such a regime has lead to various applications, e.g., 
 exploiting the well known Purcell effect \cite{Purcell1946}  has allowed for breakthroughs in quantum technologies such as low-threshold solid-state lasers \cite{Vahala2003} and  single-photon emitters~\cite{Salter2010,somaschi2016}.
These effects allow for the engineering of the spontaneous
emission rate of an emitter, by tailoring its photonic environment. Indeed, resonant electromagnetic resonators with a narrow density of states can greatly enhance the efficiency of photonic devices.
Going beyond weak coupling, the strong-coupling regime is characterized by lower losses in the system, allowing for the observation of effects such as vacuum Rabi oscillations \cite{Haroche2013}, manifested by the coherent oscillatory exchange of energy between light and matter.
Such effects are already being exploited in  second generation quantum technologies~\cite{buluta2011natural,georgescu2012quantum}. In this strong-coupling regime, however, the light-matter coupling rate, remains much lower than the bare resonance frequencies, so that both the weak and the strong coupling regime can be adequately described by the JCM.

If the quantum light–matter interaction strength reaches a non-negligible fraction of the transition frequency of the components, the system enters the so-called ultrastrong coupling (USC) regime.
In this regime, the interaction can significantly change the system properties. For
example, the ground state of the system contains non-negligible virtual photons and virtual matter excitations.
In the past decade, USC effects between light and matter has transitioned
from a theoretical idea to an experimental reality. Nowadays, this regime has been achieved in a great variety of systems and settings \cite{Kockum2019Rev,Forn-Diaz2019}.
The experimental progress in USC physics has motivated
many theoretical studies showing interesting new effects enabled or boosted by this regime \cite{Gunter2009, Forn-Diaz2010, Todorov2010, Schwartz2011, You2011, Scalari2012, Geiser2012, Kena-Cohen2013, Xiang2013a, Maissen2014, Goryachev2014, pirkkalainen2015, Baust2016, benz2016, George2016a, Forn-Diaz2017, Yoshihara2017a, Yoshihara2017, Bayer2017, Askenazi2017, Barachati2018, Yoshihara2018, Eizner2018, Paravicini-Bagliani2019, PuertasMartinez2019, Flower2019, Jeannin2019, Wang2020, Keller2020}. Several studies explore higher-order processes in the USC regime, where the number of excitations is not conserved \cite{Law2015, Kockum2017, Kockum2017a, Stassi2017a, Macri2018, DiStefano2019}, such
as multiphoton Rabi oscillations \cite{Garziano2015} and a single photon exciting multiple atoms \cite{Garziano2016,Macri2020}.

When the light-matter interaction strength increases even further, a regime where the coupling strength exceeds the resonant frequencies of the material and/or of the quantized light modes can be achieved \cite{yoshihara2017superconducting}---the deep strong coupling (DSC) regime.
One of the most interesting effects predicted in this regime is the effective decoupling between light and matter \cite{DeLiberato2014}. A striking consequence of such a counterintuitive phenomenon is that the Purcell effect is reversed and the spontaneous emission rate, usually thought to increase with the light-matter coupling strength, tends to vanish for sufficiently large couplings. 
Such a result has been predicted considering bosonic matter excitations interacting with a multi-mode optical resonator (generalized Hopfield model), and using the Coulomb gauge. Recently, a first confirmation of this prediction has been obtained using   three-dimensional  crystals of plasmonic nanoparticles \cite{Mueller2020}.

In  subsequent work \cite{garcia2015light}, this effect has also been studied considering a single superconducting qubit interacting with a multi-mode electromagnetic resonator. In circuit QED, for a qubit interacting with a superconducting waveguide, a microscopic treatment of the light-matter coupling gives rise to a diamagnetic term, analogous to the $A^2$ term of the minimal coupling Hamiltonian. Using this spin-boson Hamiltonian, it has been shown that the spontaneous emission rate of the two-level system decreases with the intensity of the $A^2$ term, without the need to be in the DSC regimes. The results in 
Refs.~\cite{DeLiberato2014,garcia2015light} suggest that the diamagnetic term plays a key role in determining the light-matter decoupling effect. 
However, some questions remains open. Indeed, the presence of the diamagnetic term is gauge relative (e.g., it disappears in the dipole gauge); moreover, the validity of the Coulomb gauge when describing truncated atomic systems has been questioned \cite{Vukics2014, DeBernardis2018,Stokes2019}. 


Recently, some gauge issues have been solved. In particular, it has been shown how to obtain the correct quantum Rabi and Dicke Hamiltonians in the Coulomb gauge  \cite{DiStefano2019a, GarzianoPRA2020,SavastaPRA2021,Settineri2021}. It has also been shown how to derive a gauge-invariant master equation and obtain gauge-invariant emission spectra \cite{salmon2021gauge}. Throughout this article, we will use these 
recent developments.
Specifically, we provide a unified picture of light emission under incoherent pumping of the QRM, from the weak to the deep strong light-matter interaction. When the light-matter coupling strength spans from the very  weak to the deep strong coupling regimes, the spectrum of the QRM, while initially quasi-harmonic, becomes strongly anharmonic at higher couplings. Then, after reaching the deep strong limit, its behavior tends back  towards harmonicity.

While in the weak coupling regime, neglecting the counter-rotating terms and using the standard quantum optics master equation typically provides accurate results,  in the USC regime this master equation fails to describe correctly the emission spectra. This problem can be partly solved by introducing the master equation in the dressed basis \cite{Beaudoin2011}, an approach which includes the interaction between the system components in the derivation of  the dissipators.
However, this powerful approach can also fail in describing the emission of the QRM in both the weak and deep strong coupling regimes. 
To solve these problems, we study the incoherent emission of the system at any coupling strength, using a dressed-state generalized master equation (GME) working for systems displaying both harmonic, quasi-harmonic, and anharmonic spectra \cite{Settineri2018}. 
Moreover, we take particular care to derive a gauge-invariant GME, to ensure the gauge invariance of the obtained emission rates and spectra \cite{salmon2021gauge}. 

In \secref{sec: Theoretical Model1}, we first provide a description of the QRM in both the Coulomb and multipolar gauges, and then  present the GME approach which we will use for all the calculations. A detailed description of the derivation is provided in the Appendices (see also Ref.~\cite{salmon2021gauge}). The numerical calculations and their analysis are presented in \secref{sec: Results}. There, we show numerically-calculated cavity and qubit photon flux emission rates and spectra as a function of the normalized light-matter coupling strength, obtained for different effective temperatures and cavity-qubit detunings.
In order to highlight the differences between our results based on a gauge-consistent GME, and some other standard methods, in  \secref{sec: Comparison with other models}, we present (i) results obtained using the JC model and the standard quantum optics master equation, and (ii) results obtained using a dressed master equation with post-trace rotating wave approximation \cite{Beaudoin2011}. Finally,
in Sec.~\ref{sec: Conclusion}, we give
our conclusions.

In addition, the main theoretical framework used to develop all the results is fully explained in the Appendices. In \appref{Gauge_issues_cavity_qubit-bath_interaction} we model the cavity- and qubit-bath interaction by using the gauge principle, while in \appref{app: generalized master equations} we provide a demonstration of the gauge invariance of the GME.  


\section{Dissipative quantum Rabi Model}\label{sec: Theoretical Model1}

In this section, we describe the open QRM, thus including the interaction of the light and matter components with their respective reservoirs. We consider models in both the  Coulomb and the multipolar  gauge (within the dipole approximation), to show equivalence and verify gauge-invariant observables.

\subsection{Quantum Rabi model in the Coulomb gauge}\label{sec: Theoretical_Model_Coulomb_gauge}
The quantum Rabi Hamiltonian in the Coulomb gauge can be written as ($\hbar = 1$)
 \cite{DiStefano2019a}
\bea
\label{eq: coulomb_gauge_rabi_hamiltonian}
\hat{\mathcal{H}}_{\rm R} &=& \omega_c \hat{a}^\dagger \hat{a} + \frac{\omega_{q}}{2} \left\{ \sz \cos \left[ 2 \eta (\hat{a} + \hat{a}^\dagger) \right] \nonumber \right. \\
&+& \left. \sy \sin \left[2 \eta ( \hat{a} + \hat{a}^\dagger ) \right] \right\}\, .
\eea
where $\hat a$  ($\hat a^\dag$) is the photon annihilation (creation) operator and $\hat \sigma_{x,y,z}$ are Pauli matrices. 
The parameters $\omega_c$ and $\omega_q$ represent  the cavity and the qubit resonance frequencies, respectively, while $\eta = g / \omega_c$ is the normalized light matter coupling
strength.


The Hamiltonian  $\hat{\mathcal{H}}_{\rm R}$ 
can be obtained starting from the light-matter Hamiltonian in the absence of interaction $\hat{\mathcal{H}}_{0} = \hat{\mathcal{H}}_{q} + \hat{\mathcal{H}}_{\rm ph}$, where $\hat{\mathcal{H}}_{q} = {\omega_q} \sz/2$ and $\hat{\mathcal{H}}_{\rm ph} = \omega_c \adop \aop$,  by applying a suitable unitary transformation (generalized minimal coupling transformation) to $\hat{\mathcal{H}}_{q}$ only \cite{DiStefano2019a,SavastaPRA2021}. Specifically, 
\be
\hat{\mathcal{H}}_{\rm R} = \hat U \hat{\mathcal{H}}_{q} \hat U^\dag + \hat{\mathcal{H}}_{\rm ph}\, ,
\ee
where
\be\label{U}
\hat {U} = \exp[i \eta (\aop + \adop)\sx ]\, .
\ee
We observe that in the Coulomb gauge, the canonical field momentum is not modified by the interaction with the matter component, i.e., $\Pi = - \varepsilon_0 \hat E$  ($\varepsilon_0$ vacuum permittivity), such that, in this framework, the electric field operator can be written as 
 \be\label{sec:Electro_coulomb}
\hat E = i \omega_c A_0 (\hat a - \hat a^{\dag})\, ,
\ee
where $A_0$ is the zero-point-fluctuation amplitude of the field coordinate.
For simplicity we assume a simple one dimensional model, but the formalism can  be easily generalized to three dimensions.

Regardless of the chosen  gauge, when the normalized coupling strength becomes significantly large, all the eigenstates become {\it dressed} by virtual excitations (owing to the presence of  the counter-rotating terms) \cite{DiStefano2017a}, and complications arise in the theoretical description \cite{Kockum2019Rev}. 
Consequently, dissipation effects, input-output relationships, and photodetection rates \cite{DiStefano2018,LeBoite2020} cannot be introduced adequately by using the standard tools of quantum optics in the usual fashion. 
For example, considering a given quantum state of the system $\hat \rho(t)$ (i.e., the density operator of the light-matter system at time $t$), the photon rate measured by a broadband point-like detector  in the resonator is different from  $\langle \hat a^\dag \hat a \rangle_t \equiv {\rm Tr}[\hat a^\dag \hat a \hat \rho(t)]$. Instead, it is proportional to the expectation value $W_c(t) = \langle \hat {\cal E}^{-} \hat {\cal E}^{+} \rangle_t \equiv {\rm Tr}[\hat {\cal E}^{-} \hat {\cal E}^{+} \hat \rho(t)]$. Here, $ \hat {\cal E}^{+}$ [$\hat {\cal E}^{-} = (\hat {\cal E}^{+})^\dag$] is proportional to the positive (negative)-frequency electric field operator, and the expectation value is taken considering quantum states calculated in the Coulomb gauge (dressed-state representation).

Specifically, the positive-frequency electric field operator is obtained from
\be\label{sec:operators_coulomb}
\hat {\cal E}^+ = i \sum_{k >j}
\mel{j}{(\aop - \adop)}{k} \dyad{j}{k}\, ,
\ee
where $|j \rangle$ are the energy eigenstates of the Hamiltonian in \eqref{eq: coulomb_gauge_rabi_hamiltonian} with eigenvalues $\omega_j$  ordered so that $k > j$ for $\omega_k > \omega_j$. 
Using the following relationship \cite{savastaNano2021}
\be\label{sec:matrix_elemet_coulomb}
\omega_c \langle j| (\hat a - \hat a^\dag)| k\rangle = 
 \omega_{kj}\langle j| (\hat a + \hat a^\dag) | k\rangle\, ,
\ee
where ($\omega_{kj} = \omega_k - \omega_j$), it is also possible to rewrite \eqref{sec:operators_coulomb} as
\be\label{sec:operators_coulomb_2}
\hat {\cal E}^+ =  i \sum_{k >j} \frac{\omega_{kj}}{\omega_c}
\mel{j}{(\aop+\adop) }{k} \dyad{j}{k}\, .
\ee
Note that \eqref{sec:matrix_elemet_coulomb} remains valid even in the presence of very strong light-matter interactions and/or optical nonlinearities.

By using the simple input-output theory \cite{Gardiner2004}, results analogous to $W_c(t)$ can be obtained for the rate $W_c^{\rm out}(t)$ of emitted photons detected by a detector placed outside the cavity \cite{Ridolfo2012,Garziano2013}. However, the output field operators can display a different dependence on $\omega_{kj}$, arising from the density of states of the output modes and from the frequency dependence of the coupling coefficient, which  (for example) depends on the mirror reflectivity in a standard microcavity. More generally, the output field operator,  in the Coulomb gauge, can be written as
\be\label{sec: Electro_Coulomb_out}
\hat {\cal E}_{\rm out}^+ = i \sum_{k >j} \alpha (\omega_{kj})
\mel{j}{(\aop + \adop) }{k} \dyad{j}{k}\, ,
\ee
where the function $\alpha(\omega)$ encodes the specific, model-dependent dependence on the frequency.
A more rigorous input-output theory can be formulated in terms of quantized quasinormal modes~\cite{PhysRevLett.122.213901,Hughes2019,PhysRevResearch.2.033456}.

Analogously, it is possible to define the field operators describing the qubit emission $W_q(t) = \langle \hat {\cal S}^- \hat {\cal S}^+ \rangle_t$, where
\be\label{sec: Qubit_Coulomb}
\hat {\cal S}^+ = i \sum_{k >j} \alpha_q (\omega_{kj})
\mel{j}{ \sigma_x }{k} \dyad{j}{k}\, .
\ee

In  the following,  we will assume both $\alpha_c(\omega_{kj}) = \omega_{kj}/\omega_{c}$ (corresponding to 
$\hat {\cal E}_{\rm out}^+ = \hat {\cal E}^+$) and $\alpha_q(\omega_{kj}) = \omega_{kj}/\omega_q$, to be linearly dependent on the transition frequencies.
A different choice will give rise to similar spectra with different relative heights of the spectral lines.
Notice that photodetection is an energy absorption process, thus it is reasonable to assume photon detection rates which tend to zero with frequencies $\omega \to 0$.
Naturally, any realistic analysis 
covering a very large frequency range should also include  dispersion in the material model. Including the latter, however, would make the study system dependent going beyond the aim of the present general framework. 

\subsection{Quantum Rabi model in the dipole gauge}\label{sec: Theoretical_Model_dipole_gauge}
The quantum Rabi Hamiltonian can also be expressed in the dipole gauge, which yields the  form
\be
\label{eq: dipole_gauge_rabi_hamiltonian}
\hat{\mathcal{H}}'_{\rm R} = \omega_c \adop \aop + \frac{\omega_q}{2} \sz - i \eta \omega_c \left(\aop -\adop\right) \sx\, + \omega_c \eta^2.
\ee
To be clear, 
by dipole gauge, we mean the multipolar gauge after the dipole approximation \cite{cohen1997photons,Wubs2004}. Light-matter Hamiltonians in the multipolar or dipole gauge can be obtained from the Coulomb gauge (minimal coupling replacement), after a unitary Power–Zienau–Woolley (PZW) transformation \cite{babiker1983derivation}.

The last term in \eqref{eq: dipole_gauge_rabi_hamiltonian} is often disregarded, since it has no dynamical consequences. In the following, when needed, as for the Hamiltonian in the dipole gauge $\hat{\mathcal{H}}'_{\rm R}$, we will use primed symbols to indicate gauge-relative quantities in the dipole gauge.
By considering a fixed polarization in the single-mode approximation, the field coordinate corresponding to the vector potential can be expressed as $\hat A = A_0 (\hat a + \hat a^\dag)$. In the dipole gauge, the field conjugate momentum is modified by the interaction with the matter system, and it is proportional to the electric displacement (induction) field 
\be
\hat \Pi' = - \hat D = -i \varepsilon_0 \omega_c A_0 (\hat a - \hat a^\dag) \, .
\ee

Thus, the electric field operator cannot be expanded in terms of photon operators only.  Indeed, due to the fact that $\hat D= \varepsilon_0 \hat E ' + \hat P$ (for a dipole in free space), 
 where $\hat P$ is the electric polarization, the electric field operator in the dipole gauge has to be expanded as
\be\label{sec: Electro_dipole_gauge}
\hat E' = i \omega_c A_0 (\hat a' - \hat a'^{\dag})\, ,
\ee
where (see Ref.~\cite{Settineri2021})
\be\label{dipolo_photon_operator}
\hat a' = \hat R \hat a \hat R^\dag = 
\hat a+i \eta \hat \sigma_x\, ,
\ee
and
\be\label{T}
\hat {R} = \hat U^\dag = \exp[-i \eta (\aop + \adop)\sx ]\, .
\ee

This unitary operator essentially implements the PZW transformation
for a truncated TLS model and in the dipole approximation \cite{SavastaPRA2021}.
Of course, the different representations provide the same energy levels for the light-matter system. They also provide identical expectation values if operators and quantum states are  both properly transformed \cite{Settineri2021}.  
We observe that the operators $\hat a'$ and $\hat a'^{\dag}$ satisfy the same commutation relations of the bosonic operators $\hat a$ and $\hat a^{\dag}$.
Moreover, since $\hat a' + \hat a'^{\dag} = \hat a + \hat a^\dag$, the vector potential can also be expressed as $\hat A = A_0 (\hat a' + \hat a'^\dag)$.
We observe that, in the dipole gauge,  $\hat a'$ and $\hat a'^{\dag}$ (instead of $\hat a$ and $\hat a^{\dag}$) describe the creation and annihilation of the field quanta, as it is clear from \eqref{sec: Electro_dipole_gauge}. 

The quantum Rabi Hamiltonians in \eqref{eq: coulomb_gauge_rabi_hamiltonian} and \eqref{eq: dipole_gauge_rabi_hamiltonian} are related by a gauge transformation, implemented by the unitary operator in \eqref{T}:
\be
\mathcal{\hat H}'_{\rm R} = \hat {R}\mathcal{\hat H}_{\rm R}\hat {R}^\dag\, .
\ee

The photon rate measured by a broadband point-like detector in the resonator can be expressed also in the dipole gauge as $W_c'(t)= \langle \hat {\cal E}'^{-} \hat {\cal E}'^{+} \rangle'_t$, where the  expectation values are calculated using the eigenstates $|j' \rangle$ of \eqref{eq: dipole_gauge_rabi_hamiltonian}. Therefore, $\hat {\cal E}'^{+}$ is now proportional to the positive-frequency electric field operator in the dipole gauge,
\be\label{eq: Electro_dipole_gauge}
\hat {\cal E}'^+ = i\sum_{k >j}
\mel{j'}{(\hat a' - \hat a'^\dag)}{k'} \dyad{j'}{k'}\, .
\ee
From \eqref{dipolo_photon_operator}, clearly we see that the photon operators are not gauge invariant. Thus, in order to obtain correct results in the dipole gauge \cite{Settineri2021}, it is essential to properly take into account how these operators change under the $\hat {U}$-transformation [see \eqref{dipolo_photon_operator}]. 

Choosing the dipole gauge, without transforming the photonic operators accordingly, can lead to an erroneous evaluation of the emitted photon rate, as shown in \secref{sec: Results}.
Specifically, an incorrect photon rate is obtained by  using $\hat a$ ($\hat a^{\dagger} $) instead of $\hat a'$ ($\hat a'^{\dagger} $)   in the dipole gauge.
In this case, the photon rate becomes
\be\label{sec:electro_rate_wrong}
{\widetilde W}_c'= \langle \hat {\cal E}_w'^{-} \hat {\cal E}_w'^{+} \rangle'\, ,
\ee
where
\be\label{Electro_dipolo_wrong}
\hat {\cal E}_w'^+ = i\sum_{k >j}
\mel{j'}{ (\hat a - \hat a^\dag) }{k'} \dyad{j'}{k'}. 
\ee
Also note, ${\widetilde W}_c' \neq {W}_c'= W_c$ (see \secref{sec: Results}).

The useful relationship shown by \eqref{sec:matrix_elemet_coulomb} can be appropriately transformed in the dipole gauge as
\be\label{sec:matrix_elemet_dipole}
\omega_c \langle j'|(\hat a' - \hat a'^\dag)| k'\rangle = 
 \omega_{kj}\langle j'| (\hat a' + \hat a'^\dag)| k'\rangle\, ,
\ee
such that, 
\be\label{sec:electro_dipolo_2}
\hat {\cal E}'^+ = i \sum_{k >j} \frac{\omega_{kj}}{\omega_c}
\mel{j'}{ (\hat a' + \hat a'^\dag )}{k'} \dyad{j'}{k'}\, ,
\ee
and we also have:
\be
\mel{j'}{ (\hat a' + \hat a'^\dag) }{k'} = \mel{j}{ (\hat a + \hat a^\dag)}{k}.
\ee

\subsection{Theoretical description of losses and quantum noise}\label{sec: Theoretical_Model_dissipation_gauge_invariant}
In order to calculate emission rates and emission 
spectra of the QRM, from the weak to the DSC regime, we describe the dissipative system dynamics considering a GME in the dressed basis \cite{Settineri2018}, 
\be\label{ME}
\dot{\hat{\rho}} = - i \left[ \hat{\mathcal{H}}_R, \hat{\rho} \right] + \mathcal{L}_{\text{g}} \hat{\rho}\, ,
\ee
where the dissipator $\mathcal{L}_{\text{g}}$ contains two contributions $\mathcal{L}_{\text{g}}=\mathcal{L}^{\rm c}_{\rm g} + \mathcal{L}^{\rm q}_{\rm g}$, arising from the cavity-bath (c) and the qubit-bath (q) interaction [see \appref{app: generalized master equations} and in particular \eqref{GME}]. It remains valid at any light-matter coupling strength.
By  using this approach, we also include the interaction of the matter and light components of the system with individual reservoirs that can be at different temperatures [see \eqref{NT}].

Starting from a gauge-invariant approach (see \appref{app: generalized master equations}), the obtained photonic and atomic decay rates associated to given system transitions can be written as
\bea\label{decay_rates}
\Gamma^{\rm c}_{kj} &=& \kappa \frac{\omega_{kj}}{\omega_c}|\langle j| \hat a+ \hat a^\dag |k \rangle|^2\, ,\nonumber\\
\Gamma^{\rm q}_{kj} &=& \gamma \frac{\omega_{kj}}{\omega_q}|\langle j| \hat \sigma_x |k \rangle|^2\, ,
\eea
where $\kappa$ and $\gamma$ are the bare (in the absence of cavity-atom coupling) loss rates for the photon and the atom, and Ohmic reservoirs have been considered (i.e., the rates scale linearly with frequency).
Since the matrix elements in \eqref{decay_rates} are gauge-invariant, then $\Gamma^{\rm c(q)}_{kj} = \Gamma^{\rm c(q)}_{k'j'}$.
As done before, we  label the quantum states and operators in the dipole gauge with the prime superscript.

These {\it gauge invariant} decay rates have been derived starting from cavity and qubit reservoir interactions obtained by invoking the gauge principle (see \appref{Gauge_issues_cavity_qubit-bath_interaction}). This also gives rise to a gauge-invariant GME (see \appref{app: generalized master equations}), which, in turns, provides gauge-invariant emission rates and spectra (see \secref{FormulasSpectra}). 


\subsection{Formulas for cavity and qubit emission spectra}
\label{FormulasSpectra}

In addition to cavity and qubit emission rates, we also present emission spectra, which allows one to obtain information on the frequency of the emitted photons and, indirectly, on the system dynamics under incoherent excitation (for example). The steady-state  cavity and qubit emission spectra (obtained from the steady-state density operator $\hat \rho_{ss}$ {by applying the quantum regression theorem \cite{Mandel1995,Gardiner2004}}) can be defined as
\bea\label{spectra_dressed}
\tilde S_c(\omega) = {\rm Re} \int_0^\infty d \tau  e^{-i \omega \tau}\langle \hat {\cal E}^-(t+ \tau) \hat {\cal E}^{+}(t) \rangle_{\rm ss} \nonumber \\
\tilde S_q(\omega) = {\rm Re} \int_0^\infty d \tau  e^{-i \omega \tau}\langle \hat {\cal S}^-(t+ \tau) \hat {\cal S}^{+}(t) \rangle_{\rm ss}\, .
\eea

Note that, the above definition is valid only when considering steady-state (ss). In the USC regime, true steady-state is achieved only under incoherent pumping. Under coherent drive, the counter-rotating terms often determine the presence of oscillations in the signals, even for times much longer than coherence times. In this case the spectra have to be defined introducing an additional time integration \cite{salmon2021gauge}.

Instead of \eqref{spectra_dressed}, we will adopt a slightly different formulas for the spectra. Actually, the frequency of photons detected after a spectrum analyzer tuned  at a frequency $\omega$ is just $\omega$ and not $\omega_{kj}$, even if they originate from a specific downward transition $| k \rangle \to |j \rangle$ (which, however, is broadened by the interaction of the system with the reservoirs). Hence, it is somewhat more accurate to replace $\omega_{kj}$ with $\omega$ in the emission spectra formulas. In this way, in the low-frequency limit, the spectra goes to zero as expected. 
However, this replacement can affect the high-frequency behavior of the spectra, especially when some cut-off mechanism is not introduced.

{By making use of Eqs.~(\ref{sec:matrix_elemet_coulomb},\ref{sec:matrix_elemet_dipole}), and by applying the replacement $\omega_{kj} \to  \omega$,  we obtain}
\bea
 S_c(\omega) &=& \frac{\omega^2}{\omega^2_c}{\rm Re} \int_0^\infty d \tau  e^{-i \omega \tau}\langle \underline{\hat {\cal E}}^-(t+ \tau) \underline{\hat {\cal E}}^{+}(t) \rangle_{\rm ss} \nonumber \\
 S_q(\omega) &=& \frac{\omega^2}{\omega^2_q} {\rm Re} \int_0^\infty d \tau  e^{-i \omega \tau}\langle \underline{\hat {\cal S}}^-(t+ \tau)  \underline{\hat {\cal S}}^{+}(t) \rangle_{\rm ss}\, ,
\label{spectra2}\eea

where 
\bea
\underline{\hat {\cal E}}^- &=&  -i \sum_{k >j} 
\mel{k}{ (\adop + \aop )}{j} \dyad{k}{j}\, \nonumber \\
\underline{\hat {\cal S}}^- &=&  -i \sum_{k >j} 
\mel{k}{ \hat \sigma_x }{j} \dyad{k}{j}\, . 
\label{ES}\eea
The results obtained using \eqref{spectra_dressed} and  \eqref{spectra2} are very similar, and small differences can emerge only on a logarithmic scale.

\section{Numerical Results}
\label{sec: Results}

In this section, we  present numerical calculations for the photon emission rates and spectra for both the cavity and the qubit, under qubit incoherent pumping, as a function of the normalized coupling strength $\eta$. The incoherent excitation of the qubit is described by coupling it with a thermal reservoir at a given effective temperature ${\cal T}_q \equiv K_{\rm B} T/ \omega_q \neq 0$ (here $K_{\rm B}$ is the Boltzmann constant).
All the results have been obtained assuming a zero temperature (${\cal T}_c =0$) cavity-reservoir, so that  the cavity emission originates from the interaction with the qubit.

The eigenstates of the quantum Rabi Hamiltonian are
obtained  by standard numerical diagonalization in a truncated,
but sufficiently large finite-dimensiona, Hilbert space. Specifically, we consider the
Hilbert space resulting from the tensor product of the qubit
basis $\{|g \rangle, \, |e \rangle\}$, and the basis constituted by the $N + 1$ photonic
Fock states up to the $N$-photon state $|N \rangle$. The truncation
number $N$ is chosen in order to ensure that the lowest $M$ energy eigenvalues and corresponding eigenvectors of interest are not appreciably modified when increasing $N$.
All the  results are obtained solving the GME in \eqref{ME}, for the density matrix of the  cavity-qubit in the dressed basis, including the lowest $M$ energy levels. The truncation number $M$ is chosen to reach convergence. Specifically, we check that the results (emission rates and spectra) do not notably change 
when increasing $M$.

In the following, where more convenient, we use a different notation for the eigenstates of the QRM, in analogy with the notation used to label the eigenstates of the JCM. At zero detuning [$\Delta \equiv (\omega_c - \omega_q)/\omega_q =0$], the excited eigenstates of the JC Hamiltonian can be written as $\vert n_{\pm} \rangle = (\vert n, g \rangle \pm \vert n-1, e \rangle)/\sqrt{2}$. 
The eigenstates of the QRM, beyond the strong coupling regime do not display the same simple structure. Here, when useful, we indicate them by generalizing the above JC notation by introducing a tilde. With this notation, the state $| \tilde 0 \rangle$ denotes the ground state, and $\vert {\tilde n}_{\pm}\rangle$ describes an eigenstate of the quantum Rabi Hamiltonian. Note that $\vert {\tilde n}_{\pm}\rangle$ tends to the corresponding JC state $\vert { n}_{\pm} \rangle$ for $\eta \ll 1$. With this notation, the energy eigenstates maintain their parity (corresponding to the parity of the integer number $\tilde n$) independently of the value of $\eta$.

All the numerical calculations involving the master equation have been obtained using $\gamma/ \omega_q = 10^{-4}$ and $\kappa /\omega_q = 10^{-3}$ and using  QuTiP under Python \cite{Johansson2012,johansson_qutip_2013}.
\begin{figure}[ht]
    \centering
    \includegraphics[width = 0.48 \textwidth]{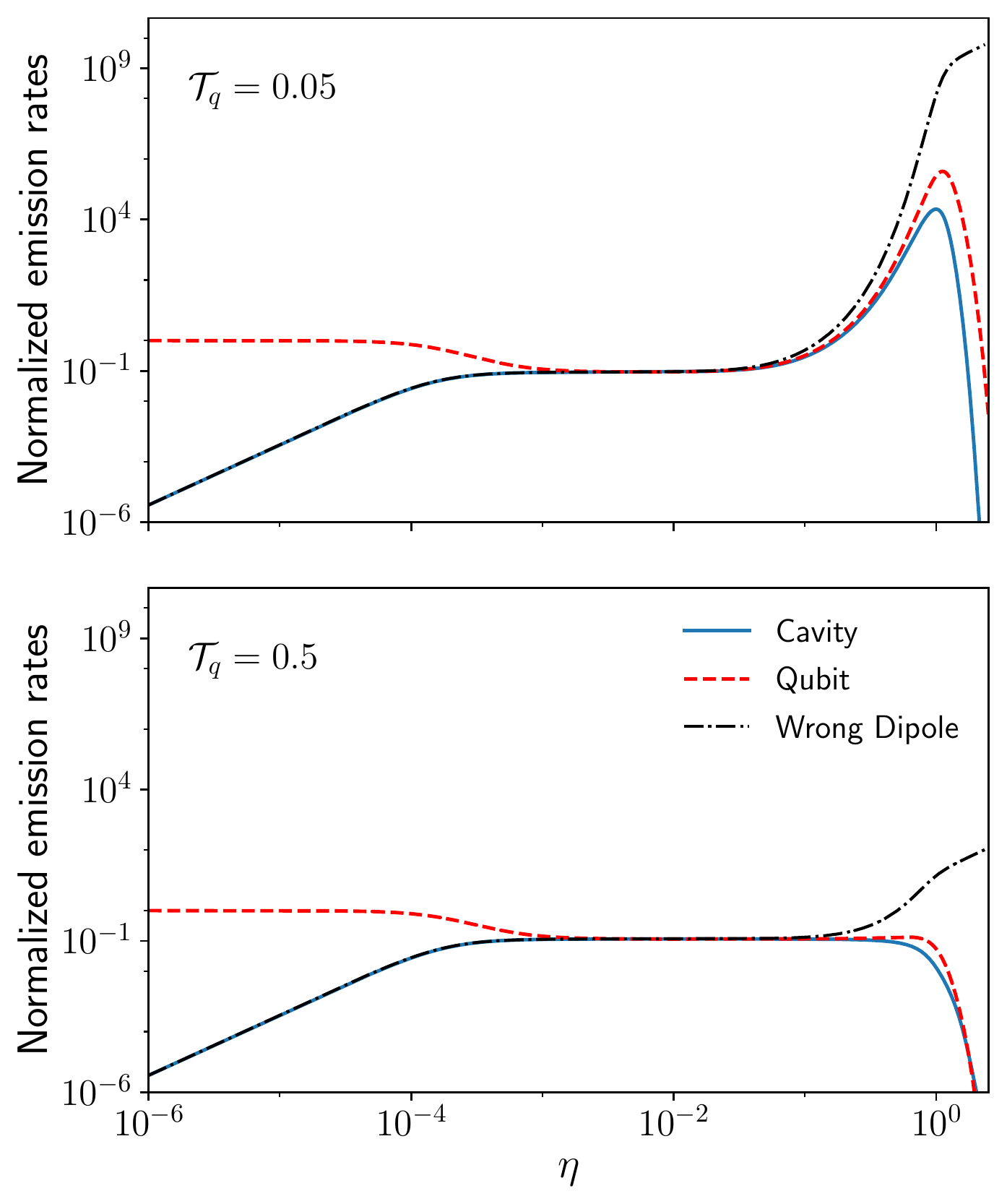}
    \caption{Numerically calculated cavity and qubit steady-state photon emission flux rates  (normalized with respect to the qubit emission rate $W^0_q$ calculated for $\eta =0$) ${\cal W}_c =W_c/W^0_q$ (blue-continuous curves), and ${\cal W}_q = W_q/W^0_q$ (red-dashed) versus the light-matter normalized coupling strength $\eta$. We used $\Delta =0$ (zero cavity-qubit detuning). Black-point-dashed curves indicate the cavity emission rates ${\cal {\widetilde W}}_c' ={\widetilde W}_c'/W^0_q$ calculated in the dipole gauge, using the {\em wrong} positive-frequency  electric field operator in \eqref{Electro_dipolo_wrong}. The upper panel (a) has been obtained for ${\cal T}_q = 5 \times 10^{-2}$, the lower one (b) for ${\cal T}_q = 5 \times 10^{-1}$.} \label{fig: ss_excitations}
\end{figure}
\begin{figure}[ht]
    \centering
    \includegraphics[width = 0.5 \textwidth]{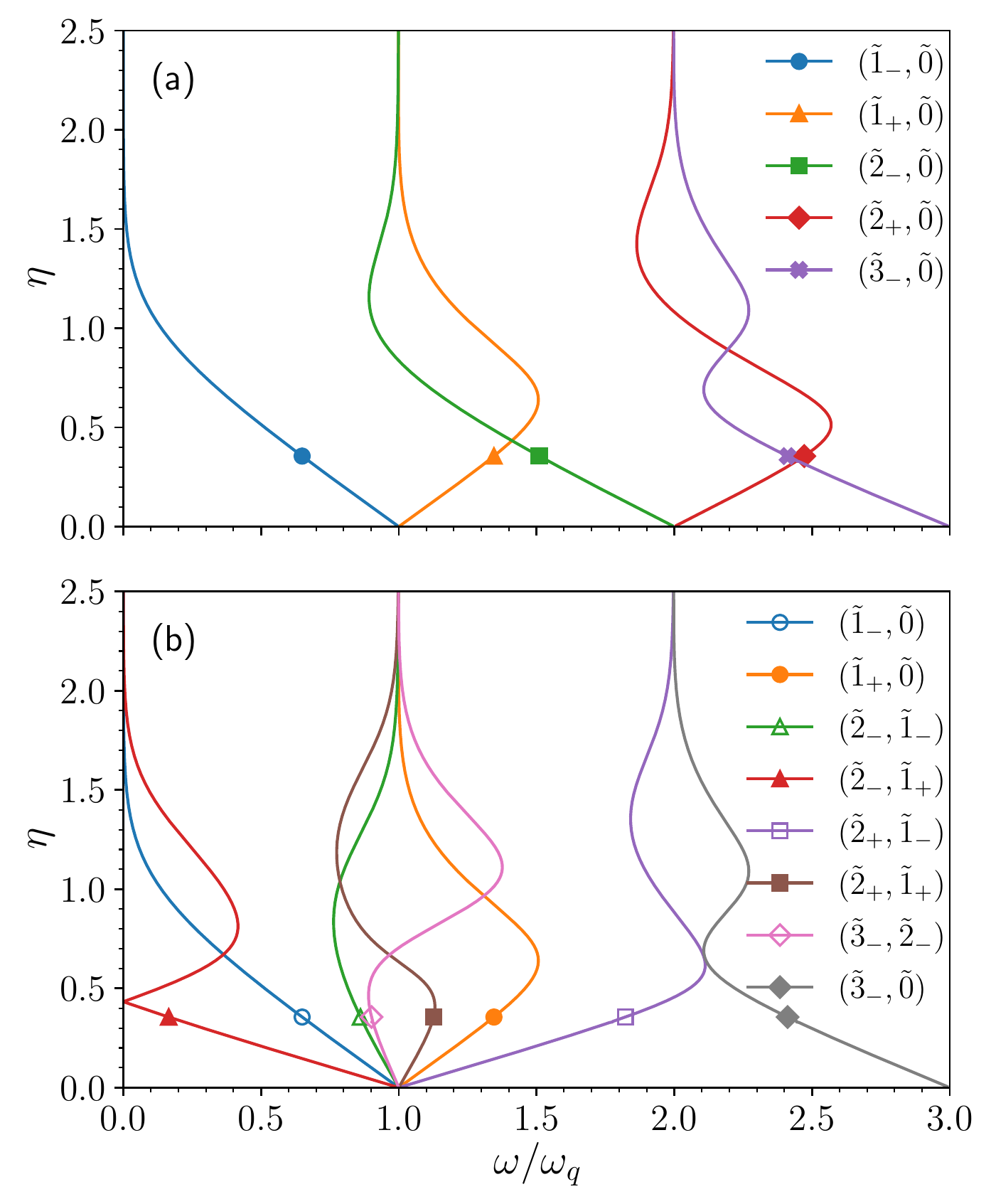}
    \caption{Normalized energy levels and transition energies versus $\eta$, for $\Delta =0$. (a) Lowest normalized energy levels (with the ground state energy as reference) $|\omega_{{\tilde j}_\pm} - \omega_{\tilde 0}|/\omega_q$ of the QRM. (b) Normalized parity-allowed transition energies $|\omega_k - \omega_j|/ \omega_q$ for the lowest eigenstates of the QRM.} \label{fig: energy transitions}
\end{figure}
\begin{figure}[ht]
    \centering
    \includegraphics[width = 0.48 \textwidth]{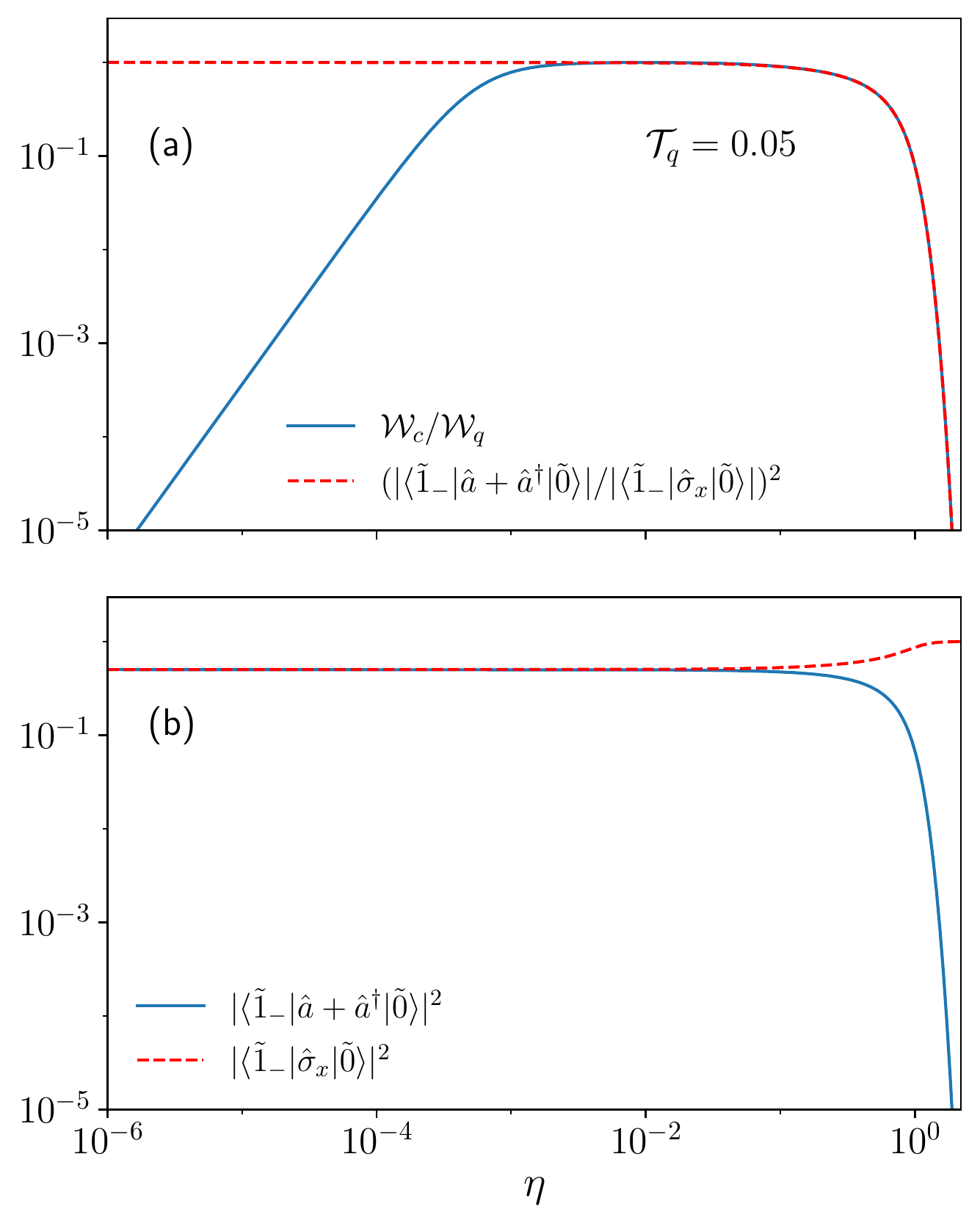}
    \caption{(a) Ratio of the cavity to the qubit emission rates ${\cal W}_c/{\cal W}_q$, for detuning $\Delta =0$, ${\cal T}_q = 5 \times 10^{-2}$, as a function of the normalized coupling strength $\eta$. The blue continuous curve describes the numerically calculated ratio, while the red-dashed one is the corresponding analytical approximated result ${\cal W}_c/{\cal W}_q \simeq |\langle \tilde 1_-|(\hat a + \hat a^\dag)| \tilde 0 \rangle|^2 / |\langle \tilde 1_-| \hat \sigma_x| \tilde 0 \rangle|^2$ [see \eqref{boh}]. The two curves tend to coincide for $\eta > 5 \times 10^{-3}$. In (b) the square modules of the matrix elements that determine the approximate ratio in (a) are shown.} \label{fig: ratio matrix element}
\end{figure}

\subsection{Zero cavity-qubit detuning}

Here, we present the results obtained analyzing emission rates and spectra at zero cavity-qubit detuning 
$\Delta \equiv (\omega_c - \omega_q) / \omega_q = 0$ 

\subsubsection{Cavity and qubit emission rates}
Figure \ref{fig: ss_excitations} shows the numerically calculated cavity and qubit steady-state photon emission flux rates (normalized with respect to the qubit emission rate $W^0_q$ for $\eta =0$) ${\cal W}_c =W_c/W^0_q$ (blue-continuous curve), and ${\cal W}_q = W_q/W^0_q$ (red-dashed) versus the light-matter normalized coupling strength $\eta$, calculated at two different effective temperatures of the qubit reservoir. The black-point-dashed curve indicates the cavity emission rates ${\cal {\widetilde W}}_c' ={\widetilde W}_c'/W^0_q$ calculated in the dipole gauge, using the {\em wrong} positive-frequency  electric field operator in \eqref{Electro_dipolo_wrong}. 
We first observe that, the methods discussed in \secref{sec: Theoretical Model1} allow  to calculate emission rates for very different values of $\eta$, within a unified theoretical framework from the weak to the DSC regimes.

Figure \ref{fig: ss_excitations} clearly shows that, at low coupling strengths (i.e., weak coupling regime), a continuous increase of the cavity-emission rate with increasing $\eta$ occurs for both, very low and higher effective temperatures (Purcell effect).
When the system approaches the SC regime [$\omega_q \eta \simeq (\kappa + \gamma)/4$] a plateau is reached, in which the emission rate is equally shared between the atom and the cavity. By increasing further the coupling beyond the onset of USC ($\eta > 0.1$), a strong enhancement of both the cavity and qubit emission can be observed at low temperature ${\cal T}_q = 5 \times 10^{-2}$ [\figpanel{fig: ss_excitations}{a}]. It originates from the decrease of the transition frequency $\omega_{\tilde 1_-,\tilde 0}$ between the lowest excited state and the ground state for increasing values of $\eta$ [see \figpanel{fig: energy transitions}{b}]. The strong decrease of $\omega_{\tilde 1_-,\tilde 0}$ enables the increase of the occupancy of the state $| \tilde 1_- \rangle$  at  very low effective temperatures. Such a population growth determines an increase in the emission rate (of photons at  frequency $\omega_{\tilde 1_-,\tilde 0}$), which can be observed in \figpanel{fig: ss_excitations}{a}. 
The same behavior  is not observed at a significantly higher temperature [\figpanel{fig: ss_excitations}{b}]. In this case, the state $| \tilde 1_- \rangle$ can already be populated at small values of $\eta$.

When increasing $\eta$ to values beyond the DSC regime  ($\eta > 1$), then both
${\cal W}_c$ and ${\cal W}_q$ decrease rapidly. This behavior can be understood by looking at the energy eigenvalues in \figpanel{fig: energy transitions}{a} in the large $\eta$ limit. Indeed, in this regime, all the transition frequencies tend to become flat, equally spaced, and 
two-by-two quasi-degenerate. 
Theoretical analysis \cite{Settineri2021} shows that, beyond the DSC regime, the quasi-degenerate light-matter  eigenstates tend to factorize as $| n, g \rangle$  and $|n,e \rangle$, so that the system tends to  decouple from the qubit reservoir ($\Gamma^{\rm q}_{kj} \to 0$) and cannot be significantly excited. Specifically, the matrix elements $\langle j | \hat \sigma_x | k \rangle$ are different from zero only for states such that $\omega_{kj} \to 0$. As a consequence, $\Gamma^{\rm q}_{kj} \to 0$.
\begin{figure}[ht]
    \centering
    \includegraphics[width = 0.48 \textwidth]{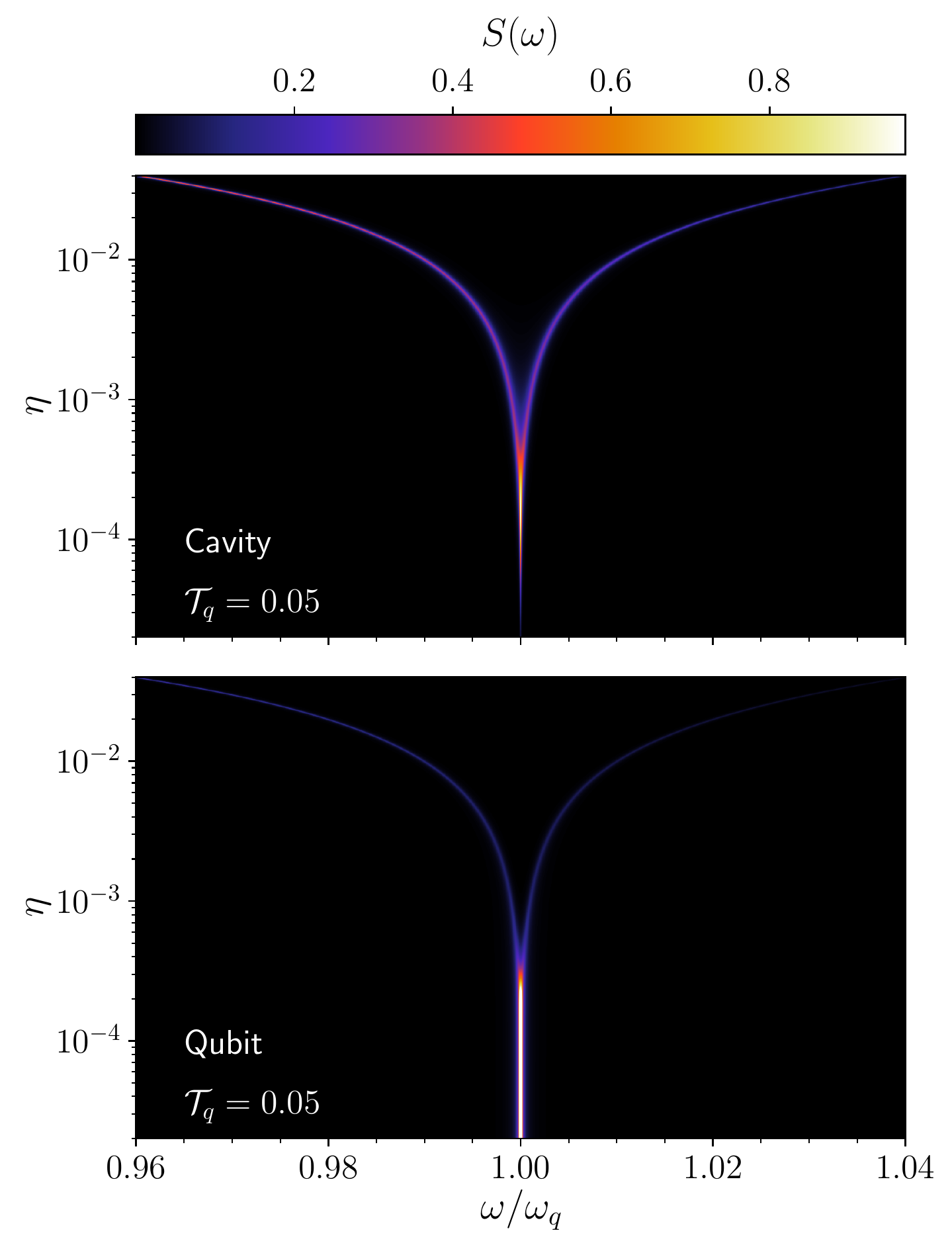}
    \caption{Cavity $S_c(\omega)$ and qubit $S_q(\omega)$ emission spectra in the weak and strong coupling regimes, calculated for $2 \times 10^{-5} < \eta < 4 \times 10^{-2}$, and for $\Delta =0$. The spectra have been obtained under weak incoherent excitation of the qubit. We used an effective qubit temperature ${\cal T}_q = 5 \times 10^{-2}$. The spectra have been normalized, so that the highest peak in each density plot is set at $1$. The signal above the horizontal line in the lower panel (qubit emission spectra) has been magnified by a factor $10$. } \label{fig: spectrum_weak-strong}
\end{figure}
\begin{figure*}[ht]
    \centering
    \includegraphics[width = \textwidth]{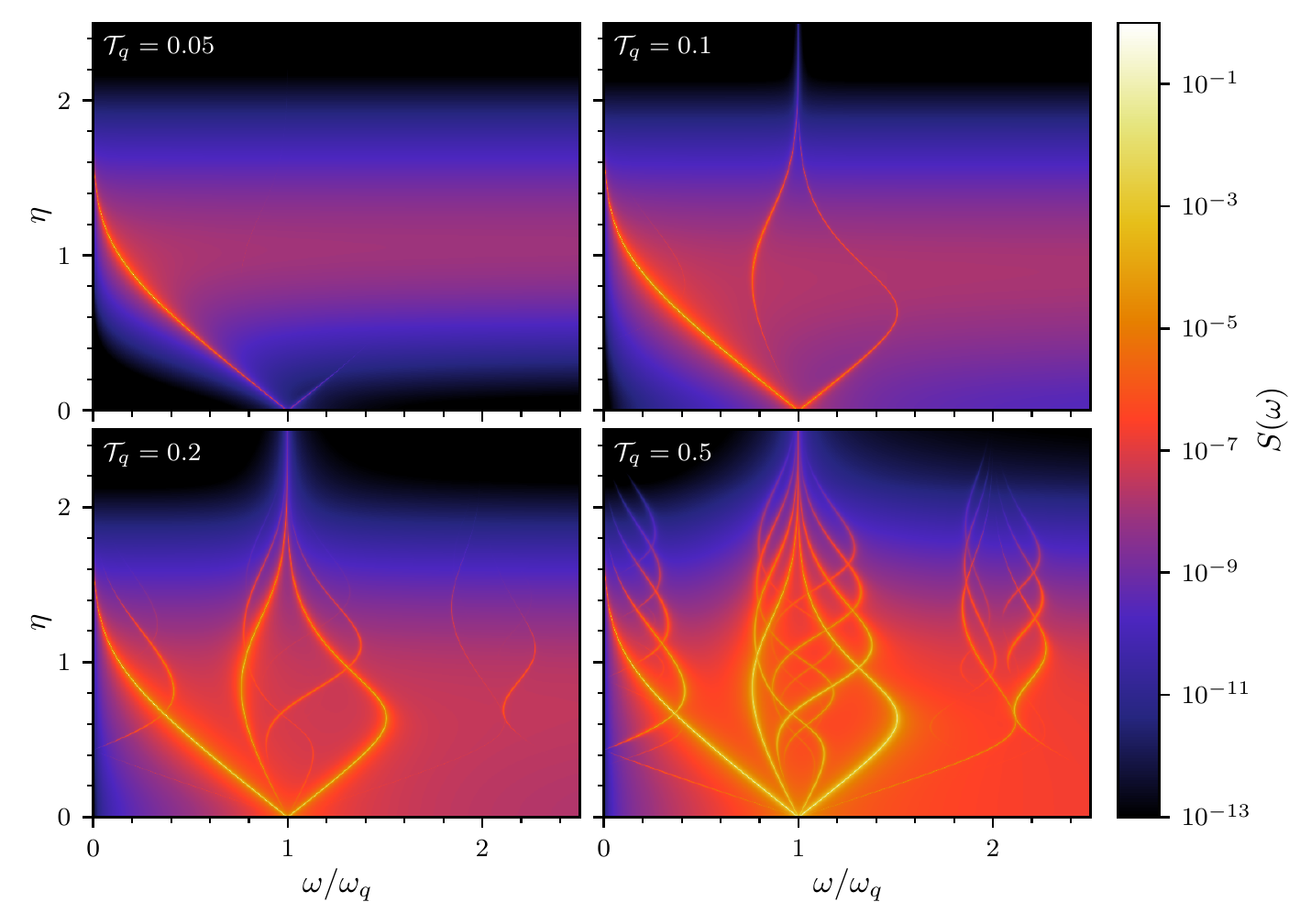}
    \caption{Logarithmic 2D plots of cavity emission spectra $S_c(\omega)$ for values of $\eta$ reaching the USC and DSC regimes, obtained using four different effective qubit temperatures ${\cal T}_q$, and $\Delta =0$. The spectra have been normalized, so that the highest peak in each density plot is set at $1$. Increasing the temperature, additional lines originating from transitions involving higher energy levels appear. Most of the emission lines correspond to transition energies shown in \figref{fig: energy transitions}.} \label{fig: spectrum_strong-deep}
\end{figure*}

We also notice that in the large $\eta$ limit, especially at low temperatures [see \figpanel{fig: ss_excitations}{a}], the cavity emission rate ${\cal W}_c$ goes to zero more rapidly than ${\cal W}_q$. 
This feature is shown more clearly in \figpanel{fig: ratio matrix element}{a}, where the ratio ${\cal W}_c/{\cal W}_q$ is displayed (blue-continuous curve). The figure shows the standard Purcell-like dependence in the weak-coupling regime, while ${\cal W}_c$ saturates to $0.5$ in the strong coupling and up to the onset of the USC regime. Still increasing $\eta$, the ratio ${\cal W}_c/{\cal W}_q$ then decreases dramatically: the Purcell effect is reversed. A similar behavior has been predicted  in a polariton system arising from the interaction of photons in a multi-mode cavity with collective electronic (bosonic) excitations \cite{DeLiberato2014}, and experimentally confirmed in three-dimensional crystals of plasmonic nanoparticles interacting with light in the DSC regime \cite{Mueller2020}.

It is interesting to explore the impact on the photon emission rate of not taking into account the proper transformation of the photon operators, when adopting the dipole gauge (${\cal {\widetilde W}}_c'$). The dot-dashed black curves in \figref{fig: ss_excitations} displays such a wrong result. The differences with respect to  the correct results ${\cal{W}}_c$ become evident at the onset of the USC regime ($\eta \sim 0.1$). In the DSC regime, the impact of this mistake becomes very relevant: ${\cal{W}}_c'$ does not go to zero and becomes orders of magnitudes larger than the qubit emission rate ${\cal{W}}_q$.

In \figpanel{fig: ratio matrix element}{a} it is also displayed an approximate analytical derivation of the ratio ${\cal W}_c/{\cal W}_q$.
When the coupling rate $\eta$ is strong enough to split sufficiently the two lowest energy excited levels (so that, at very low effective temperatures, only the system ground state and the first excited level are populated) the higher energy levels can be neglected (effective dressed two-level system). For these values of $\eta$, the cavity and emission rates can be simplified to
\bea
W_c &\simeq& \frac{\omega^2_{\tilde 1_-, \tilde 0}}{\omega^2_c} |\langle \tilde 1_- |(\hat a + \hat a^\dag)| \tilde 0 \rangle|^2\, \rho^{\rm ss}_{\tilde 1_-, \tilde 1_-}\, , \nonumber \\
W_q &\simeq& \frac{\omega^2_{\tilde 1_-, \tilde 0}}{\omega^2_q} |\langle \tilde 1_- | \hat \sigma_x | \tilde 0 \rangle|^2\, \rho^{\rm ss}_{\tilde 1_-, \tilde 1_-}\, ,
\label{boh}\eea
where $\rho^{\rm ss}$ indicates the steady state density operator. It is interesting to note that the ratio ${\cal W}_c/{\cal W}_q $ in \eqref{boh}
is independent of the density matrix. This approximate value of the ratio (red-dashed curve), shown in \figpanel{fig: ratio matrix element}{a}, is able to reproduce accurately the numerically calculated values of ${\cal W}_c/{\cal W}_q$ for $\eta > 10^{-3}$.
In \figpanel{fig: ratio matrix element}{b} it is shown the square modules of the transition matrix elements  $|\langle \tilde 1_- |(\hat a + \hat a^\dag)| \tilde 0 \rangle|^2$ and $|\langle \tilde 1_- | \hat \sigma_x | \tilde 0 \rangle|^2$ as a function of the normalized coupling $\eta$. In the high $\eta$ limit, the first one goes to $0$ and the second to $1$. 

In summary, the scenario in the high $\eta$ and low qubit-temperature limit is the following: the qubit spontaneous emission rate ${\cal W}_q$ goes to zero because $\Gamma^{\rm q}_{\tilde 1_-,\tilde 0} \to 0$ (qubit decoupling from its reservoir) and the cavity emission rate ${\cal W}_c$ goes to zero even more rapidly (at increasing $\eta$) because $|\langle \tilde 1_- |(\hat a + \hat a^\dag)| \tilde 0 \rangle|^2 \to 0$ and  $|\langle \tilde 1_- | \hat \sigma_x | \tilde 0 \rangle|^2 \to 1$ (light-matter decoupling).
\begin{figure}[ht]
    \centering
    \includegraphics[width = 0.48 \textwidth]{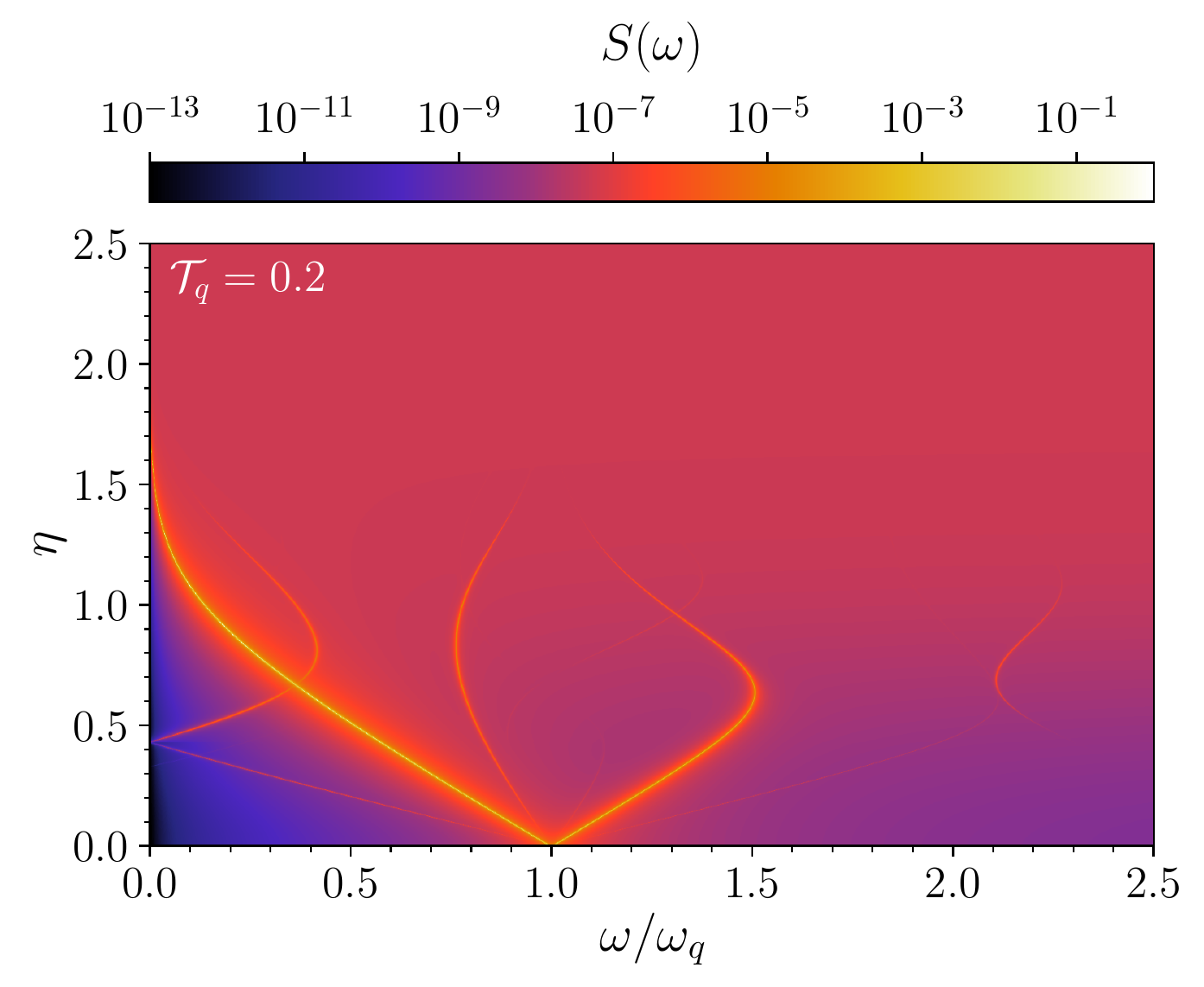}
    \caption{Logarithmic 2D plots of qubit emission spectra $S_q(\omega)$ for values of $\eta$ reaching the USC and DSC regimes, obtained at ${\cal T}_q = 0.2$, and $\Delta = 0$. The spectra have been normalized, so that the highest peak is set at $1$. The visible lines  correspond to transition energies shown in \figref{fig: energy transitions}. The origin of the flat (red) signal background for $\eta > 1.5$ is explained in the text [see \eqref{backg}].} \label{fig: spectrum_strong_deep_qubit}
\end{figure}

\subsubsection{Cavity and qubit emission spectra}

Figure~\ref{fig: spectrum_weak-strong} shows the cavity and qubit emission spectra under incoherent weak excitation of the qubit (${\cal T}_q = 5 \times 10^{-2}$) in the weak and strong coupling regimes ($2 \times 10^{-5} < \eta < 4 \times 10^{-2}$).
All the presented spectra are individually normalized, so that, in each spectral image the highest value is set to one. 
The transition from the weak (a single spectral line) to the strong (split lines)
is clearly visible in the upper panel of \figref{fig: spectrum_weak-strong}. The two lines correspond to the transitions $|\tilde 1_{\pm} \to | \tilde 0 \rangle$ indicated as $(\tilde 1_-, \tilde 0)$ and $(\tilde 1_+, \tilde 0)$ in \figref{fig: energy transitions}. In the weak coupling regime, the emission line becomes brighter at increasing values of $\eta$. In the strong coupling regime, when the two lines are sufficiently split, an asymmetry in their relative intensity can be observed. This is a direct consequence of the higher population of the lower-energy excited state  $| \tilde 1_- \rangle$ with respect to the higher-energy state $| \tilde 1_+ \rangle$ at ${\cal T}_q = 5 \times 10^{-2}$.
This behavior cannot be reproduced using the standard quantum-optics master equation where the reservoir occupations are calculated at the bare (in the absence of light-matter interaction) transition frequencies (see Sect. \ref{sec: Comparison with other models}). Across the transition from the weak to the strong coupling regime, the qubit emission (lower panel) decreases approximately by an order of magnitude (see also \figref{fig: ss_excitations}), hence, to visualize the split lines, the signal above the horizontal line in the lower panel has been magnified by a factor $10$. 

\begin{figure*}[ht]
    \centering
    \includegraphics[width = 0.8\textwidth]{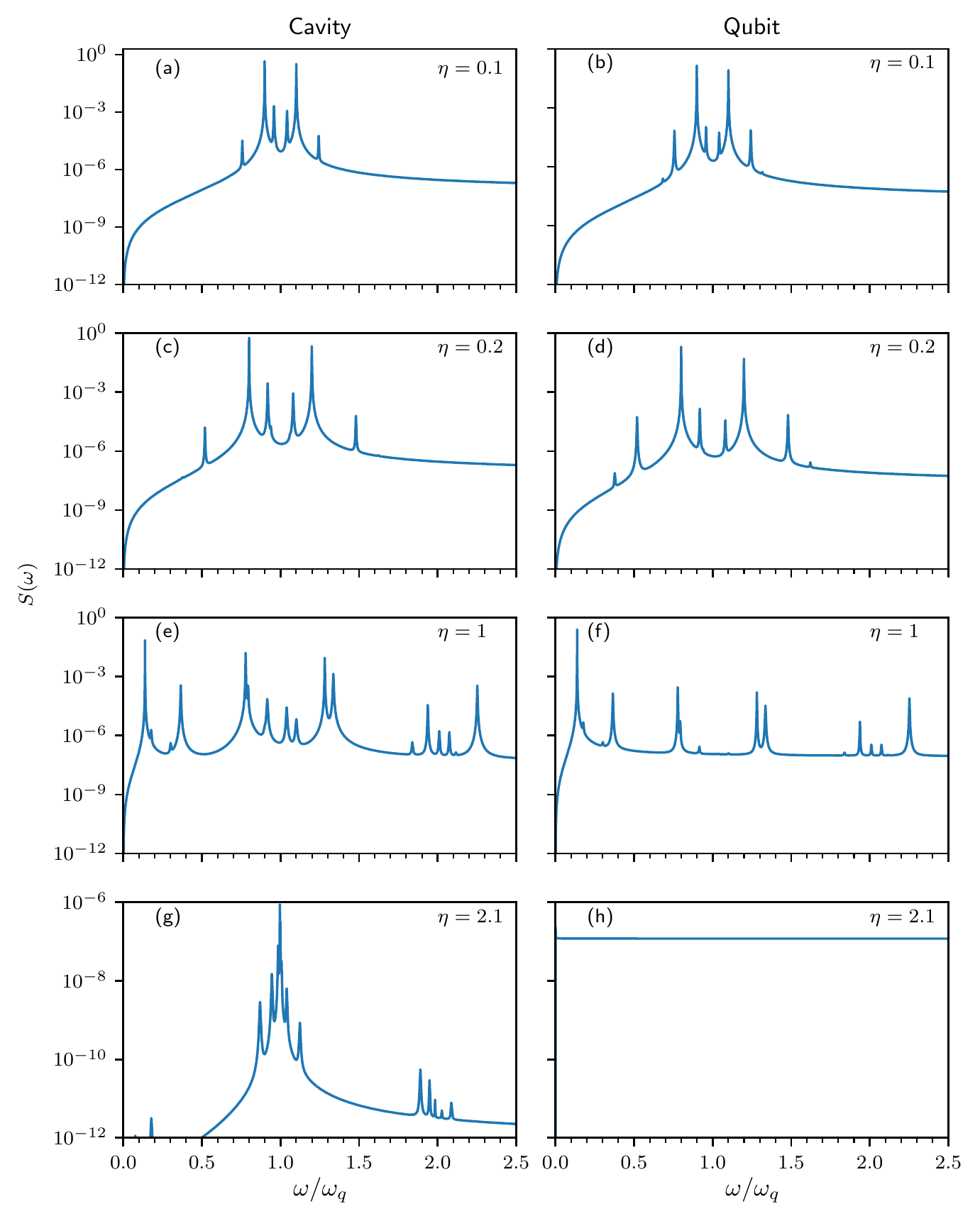}
     \caption{Cavity $S_c(\omega)$ and qubit $S_q(\omega)$ emission spectra (logarithmic), obtained at  four different values of $\eta$, increasing from the top to the bottom (detuning $\Delta = 0$).
    The cavity spectra are on the left, while the qubit ones on the right side.
    We used an effective temperature for the qubit reservoir ${\cal T}_q = 0.5$. The cavity and qubit spectra have been normalized to the highest peak which appears in panel (c) and (d), respectively. The plots show the evolution of the cavity and qubit emission spectra when increasing the normalized coupling strength $\eta$. Panel (h) reveals the flat background  appearing in the qubit emission spectra  in the DSC regime.}
    \label{fig: various_spectrum}
\end{figure*}

Figure \ref{fig: spectrum_strong-deep} displays logarithmic cavity emission spectra $\ln{[S_c(\omega)]}$ as a function of the normalized coupling strength $\eta$, calculated for four different temperatures, showing the evolution of the emission spectra from the strong to the DSC regimes. 
In contrast to the case of light-matter systems described by a harmonic Hamiltonian (see, e.g., Ref. \cite{Gambino2014}), in the present highly anharmonic case, the spectra become very rich, if the system is adequately excited. 
On this scale, the weak coupling regime (already shown in \figref{fig: spectrum_weak-strong}) is confined to a negligible portion of the $y$ axis and is not visible. At very low temperature (${\cal T}_q = 5 \times 10^{-2}$), only two spectral lines emerge, corresponding to the transitions  $(\tilde 1_\pm, \tilde 0)$ [see \figpanel{fig: energy transitions}{b}].

Notice that the transition $(\tilde 1_+, \tilde 1_-)$ is forbidden owing to parity symmetry \cite{Kockum2019Rev}.
As expected, at such a low temperature, the emission from the lowest excited level at frequency $\omega_{\tilde 1_-,\tilde 0}$ dominates. Moreover, the line at $\omega_{\tilde 1_+,\tilde 0}$ is visible only for $\eta \lesssim 0.3$. When $\eta$ increases (up to $\eta \simeq 1$), the  intensity of the line $\omega_{\tilde 1_-,\tilde 0}$ increases, due to the lowering of the ratio $\omega_{\tilde 1_-,\tilde 0}/(\omega_q {\cal T}_q)$ which causes an increase of the excited state population $\rho_{\tilde 1_-,\tilde 1_-}$. Then, for $\eta \gtrsim 1$, the population starts decreasing as a consequence of both light-matter and qubit-reservoir decoupling. This behavior is in agreement with the corresponding emission rate in \figpanel{fig: ss_excitations}{b}.

At ${\cal T} = 0.1$, the transition at frequency $\omega_{\tilde 1_+,\tilde 0}$ becomes visible for all the values of $\eta$ in the plot, although in the DSC regime, it tends to dissolve, owing to the qubit-reservoir decoupling which prevents the excitation of the excited energy levels. We also notice that a new resonance line at frequency $\omega_{\tilde 2_-,\tilde 1_-}$ appears [see \figpanel{fig: energy transitions}{b}].
\begin{figure}[htb]
    \centering
    \includegraphics[width = 0.48 \textwidth]{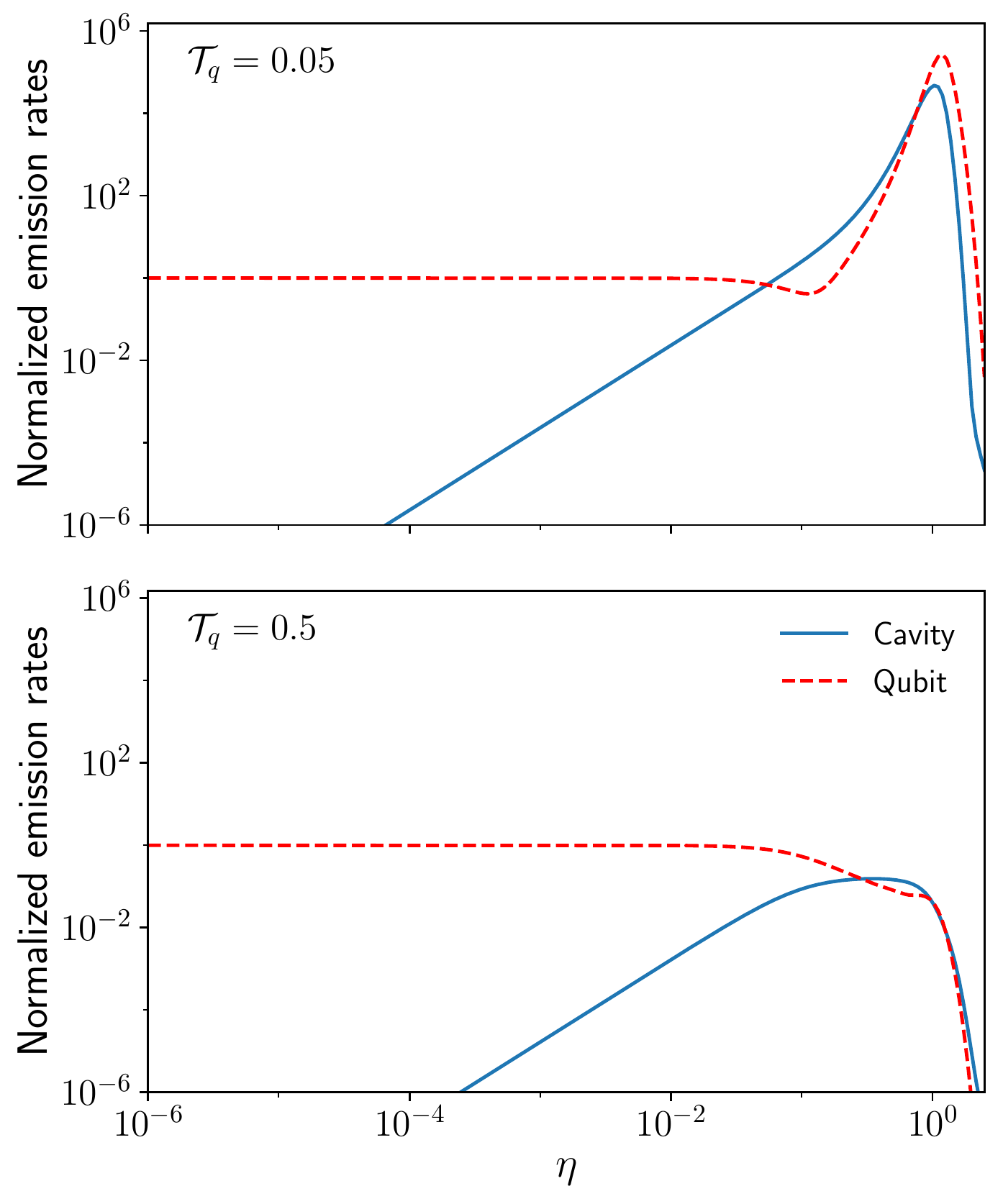}
    \caption{Cavity and qubit emission rates (normalized with respect to the qubit emission rate $W^0_q$ calculated for $\eta =0$) ${\cal W}_c =W_c/W^0_q$ (blue-continuous curve), and ${\cal W}_q = W_q/W^0_q$ (red-dashed) versus the light-matter normalized coupling strength $\eta$. We used $\Delta =-0.3$, corresponding to $\omega_c/\omega_q = 0.7$. The upper panel has been obtained for ${\cal T}_q = 5 \times 10^{-2}$, the lower one for ${\cal T}_q = 5 \times 10^{-1}$.}
    \label{fig: ss_rates_detuned_1}
\end{figure}
\begin{figure}[htb]
    \centering
    \includegraphics[width = 0.48 \textwidth]{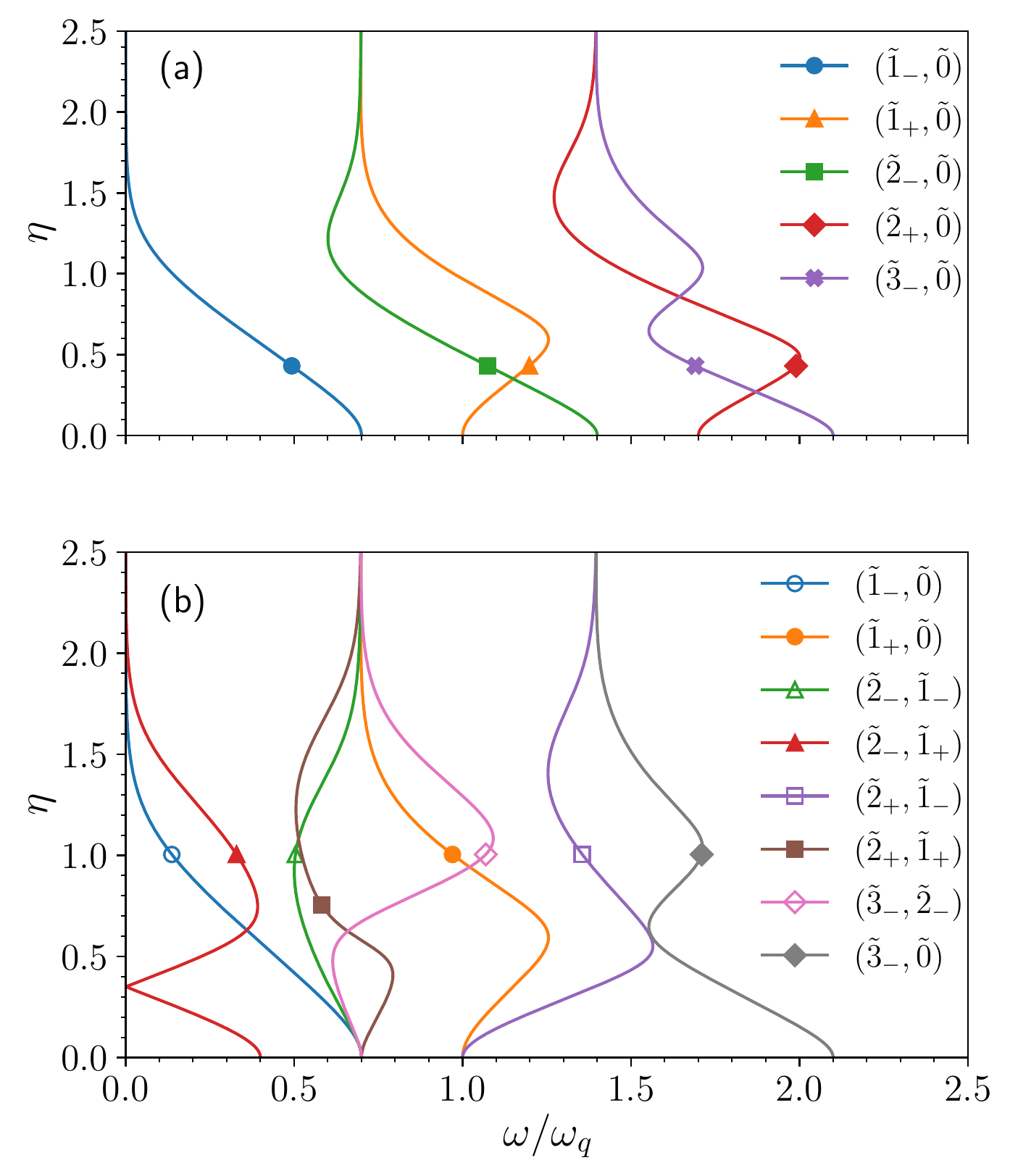}
    \caption{Normalized energy levels and transition energies versus $\eta$, for $\Delta = -0.3$ ($\omega_c / \omega_q = 0.7$). (a) lowest normalized energy levels (with the ground state energy as reference) $(\omega_{{\tilde j}_\pm} - \omega_{\tilde 0})/ \omega_q$ of the QRM. (b) Normalized parity-allowed transition energies $|\omega_{{\tilde j}_\pm} - \omega_{{\tilde k}_\pm}|/ \omega_q$ for the lowest eigenstates of the QRM.} \label{fig: energy_transitions_detuned_1}
\end{figure}
\begin{figure}[ht]
    \centering
    \includegraphics[width = 0.48 \textwidth]{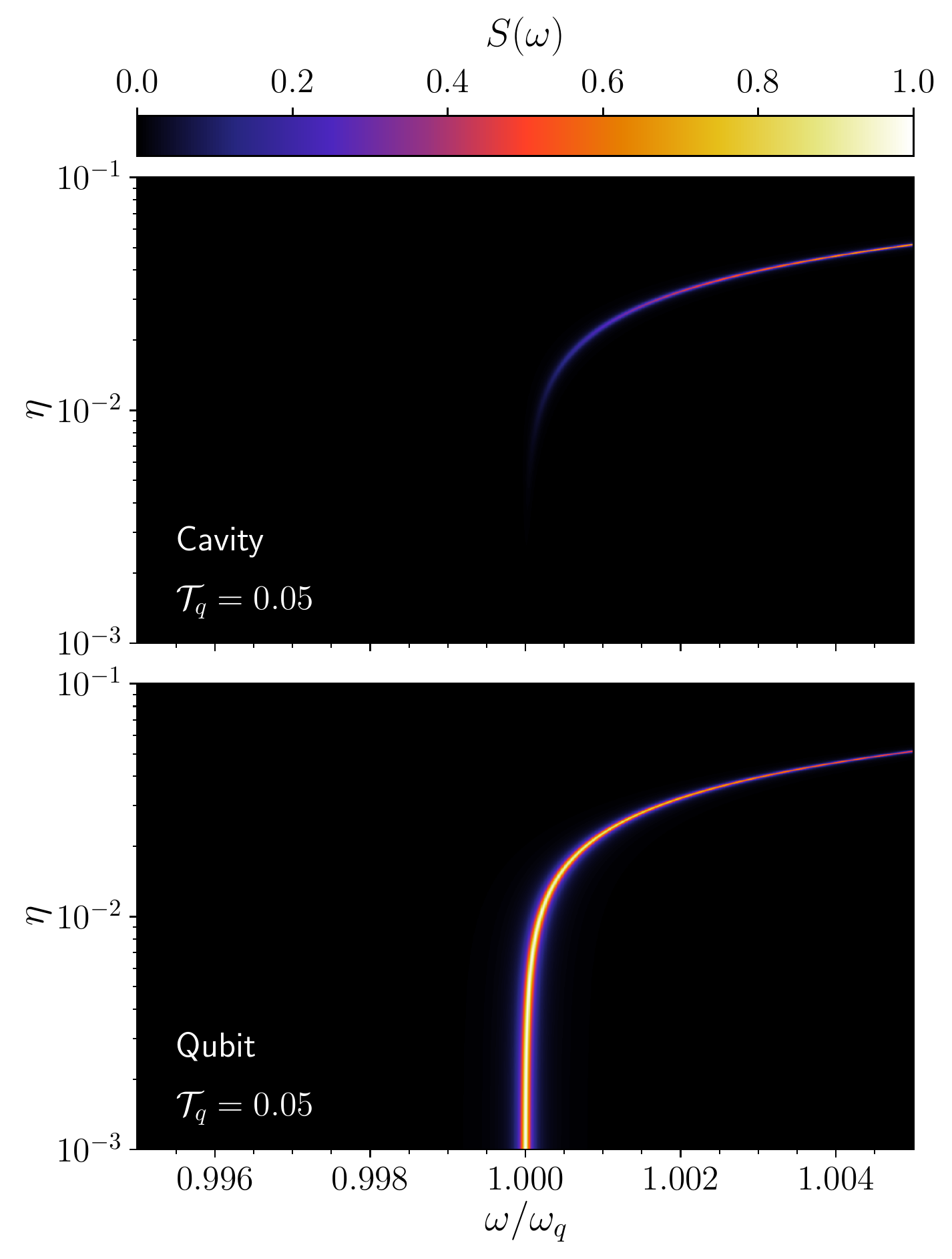}
    \caption{Cavity $S_c(\omega)$ and qubit $S_q(\omega)$ emission spectra in the weak and strong coupling regime, calculated for $10^{-3} < \eta < 10^{-1}$, and for $\Delta = - 0.3$ ($\omega_c / \omega_q = 0.7$). The spectra have been obtained under weak incoherent excitation of the qubit. We used an effective qubit temperature ${\cal T}_q = 5 \times 10^{-2}$. The spectra have been normalized, so that the highest peak in each density plot is set at $1$.}
    \label{fig: spectrum_weak_strong_detuned_1}
\end{figure}

When further increasing the temperature (${\cal T}_q = 0.2$), additional energy levels get populated and additional spectral lines appear. Most of these correspond to transitions indicated in \figpanel{fig: energy transitions}{b}. In the low-frequency range, in addition to the transition $(\tilde 1_-, \tilde 0)$, a new spectral line at $|\omega_{\tilde 2_-,\tilde 1_-}|$ appears. This transition is forbidden in the JCM, since $\langle 1_+ |(\hat a + \hat a^\dag)| 2_- \rangle =0$, at $\Delta =0$.  The crossing between the energy levels $\omega_{\tilde 2_-}$ and $\omega_{\tilde 1_-}$, occurring at $\eta \sim \bar \eta = 0.43$,  manifests as a low spectral line approaching $\omega = 0$ as $\eta \to \bar \eta$, and then (after the crossing), moving away from $\omega = 0$. At higher frequencies ($\omega/ \omega_q  \sim 1$), other two crossing spectral lines become clearly visible. As shown in \figpanel{fig: energy transitions}{b}, they correspond to the transitions $(\tilde 2_+, \tilde 1_+)$ and $(\tilde 3_-, \tilde 1_-)$, both forbidden in the JCM at zero detuning. Still at higher frequencies, other two lines are observable for $\eta \gtrsim 0.4$: they correspond to the transitions $(\tilde 2_+, \tilde 1_-)$ and $(\tilde 3_-, \tilde 0)$ [see \figpanel{fig: energy transitions}{b}]. In the latter, the involved states differ by a number of excitations $\Delta \tilde n =3$. This transition is enabled by the presence of the counter-rotating terms in the QRM \eqref{eq: dipole_gauge_rabi_hamiltonian} and represents a clear example of USC physics \cite{Kockum2019Rev}, beyond the JCM. The spectra obtained at ${\cal T}_q = 0.5$ display still richer structures with the appearance of additional lines originated by higher energy levels that get populated at this effective temperature.

The qubit emission spectra $S_q(\omega)$ calculated at ${\cal T}_q = 0.2$ are shown in \figref{fig: spectrum_strong_deep_qubit}. As expected, emission lines corresponding to those observed for $S_c(\omega)$ at the same temperature are shown. However, several differences emerge. In particular the spectral lines as a function of $\eta$ display different relative intensities. For example, (i) two of the four lines around $\omega \sim 1$  in $S_c(\omega)$ are not visible in $S_q(\omega)$; (ii) the line corresponding to the transition $(\tilde 2_-, \tilde 1_+)$ is more visible at small values of $\eta$; (iii) at increasing values of $\eta$, $S_q(\omega)$, in contrast to $S_c(\omega)$, exhibits an increasing background emission and a faster dissolving of the spectral lines at increasing values of $\eta$. All these differences  originate from the different matrix elements in \eqref{ES}  [$\langle j |(\hat a + \hat a^\dag)|k \rangle$ and $\langle j | \hat \sigma_x |k \rangle$] entering $S_c(\omega)$ and $S_q(\omega)$, respectively. 
In particular, the most peculiar feature (iii) is a direct consequence of the fact that, for $\eta \gtrsim 1.5$, the  matrix elements $\langle j | \hat \sigma_x | k \rangle \to 0$ for transitions with $\omega_{kj}$ which are significantly different from zero. Therefore, for large values of $\eta$, the qubit spectra $S_q(\omega)$ [see \eqref{spectra2}] can be approximated by a Lorentzian line-shape (times $\omega^2$) centered at $\omega \sim 0$:
\be\label{backg}
S_q(\omega) \propto \omega^2  \frac{\gamma^2}{\omega^2 + \gamma^2}\, ,
\ee
which, for $\omega \gg \gamma$, provides an almost constant background. Such behavior can be clearly observed in \figpanel{fig: various_spectrum}{h}. Notice that this high-frequency behavior can be an artifact originating from the assumption in \eqref{spectra2} and  from the absence of any cut-off mechanism. Of course,  a realistic
 behaviour in a so wide spectral range should include the specific frequency dispersion of the resonator materials, which in general is system dependent, and goes beyond the present general treatment.

Figure \ref{fig: various_spectrum} displays cavity  and qubit  emission spectra $S_{c(q)}(\omega)$ for four values of the normalized coupling strength $\eta$, calculated at ${\cal T}_q = 0.5$. The plots on the left correspond to  horizontal line-cuts of the bottom-right density plot in \figref{fig: spectrum_strong-deep}. For a more accurate comparison, the spectra in \figref{fig: various_spectrum} are reported all on the same $x$ and $y$ scale and have been normalized by the same amount, so that the highest peak in the figure is set to $1$.
Each spectral line in \figref{fig: various_spectrum} originates from a specific transition between pairs of energy levels of the QRM [see \figref{fig: energy transitions}]. 

These spectra show more in detail several features already present in the density plots. In particular, the six peaks in 
\figpanel{fig: various_spectrum}{a,b}, originate from the following transitions (from the left): $(\tilde 2_-, \tilde 1_+)$, $(\tilde 1_-, \tilde 0)$, $(\tilde 2_-, \tilde 1_-)$, $(\tilde 2_+, \tilde 1_+)$, $(\tilde 1_+, \tilde 0)$, $(\tilde 2_+, \tilde 1_-)$. In  \figpanel{fig: various_spectrum}{e,f} at $\eta = 1$, the highest peak for both the cavity and qubit spectrum is the one at the lowest frequency and corresponds to the transition $(\tilde 1_-, \tilde 0)$. Moreover, at this coupling strength, higher energy peaks around $\omega/ \omega_q \sim 2$, due to transitions from states differing by two excitations $\Delta \tilde n = 2$, can be observed. At $\eta =2$, well in the DSC regime, the intensity of the emission spectra decreases. Now $S_{c}(\omega)$ shows bunches of peaks centered at $\omega \sim 0, 1, 2$, due to the tendency of the energy levels of the system towards an harmonic spectrum in the large $\eta$ limit (see \figref{fig: energy transitions}). In \figpanel{fig: various_spectrum}{h} it is displayed a very different behavior consisting of the constant emission background explained above [see \eqref{backg}].

\subsection{Cavity-qubit interaction in the presence of detuning}

We now present  numerically calculated emission rates and spectra obtained in the case of significant qubit-cavity detuning
$\Delta \equiv (\omega_c - \omega_q) / \omega_q$, in normalized units. 
In particular, we consider the cases of $\Delta = \pm\, 0.3$.

\subsubsection{Cavity and qubit emission rates and spectra at negative detuning ($\Delta = -0.3$)}
We start from a detuning value of  $\Delta = -0.3$.
Figure \ref{fig: ss_rates_detuned_1} shows the normalized cavity and qubit emission rates, obtained with the same parameters used to calculate the results in \figref{fig: ss_excitations}, except the finite detuning. At very low effective temperature $\mathcal{T}_q = 5 \times 10^{-2}$, and at low coupling strengths, the cavity emission rate $\mathcal{W}_c$ increases linearly (on this log-log scale) as the zero-detuning case, but it is some order of magnitude smaller. As expected, the large detuning significantly reduces the energy transfer from the qubit to the cavity for $\omega_q \eta \ll \Delta$. However, the cavity emission rate becomes of the same order of magnitude of the qubit emission at $\eta \sim 0.05$, when $\omega_q \eta < \Delta$.

Moreover, the plateau in the strong coupling regime is no longer present, hence the standard Purcell behavior continues in the USC regime.
In the $\eta$ region $5 \times 10^{-2} \lesssim \eta \lesssim 0.5$, the cavity emission rate exceeds that of the qubit ($\mathcal{W}_c > \mathcal{W}_q$).
\begin{figure}[htb]
    \centering
    \includegraphics[width = 0.48 \textwidth]{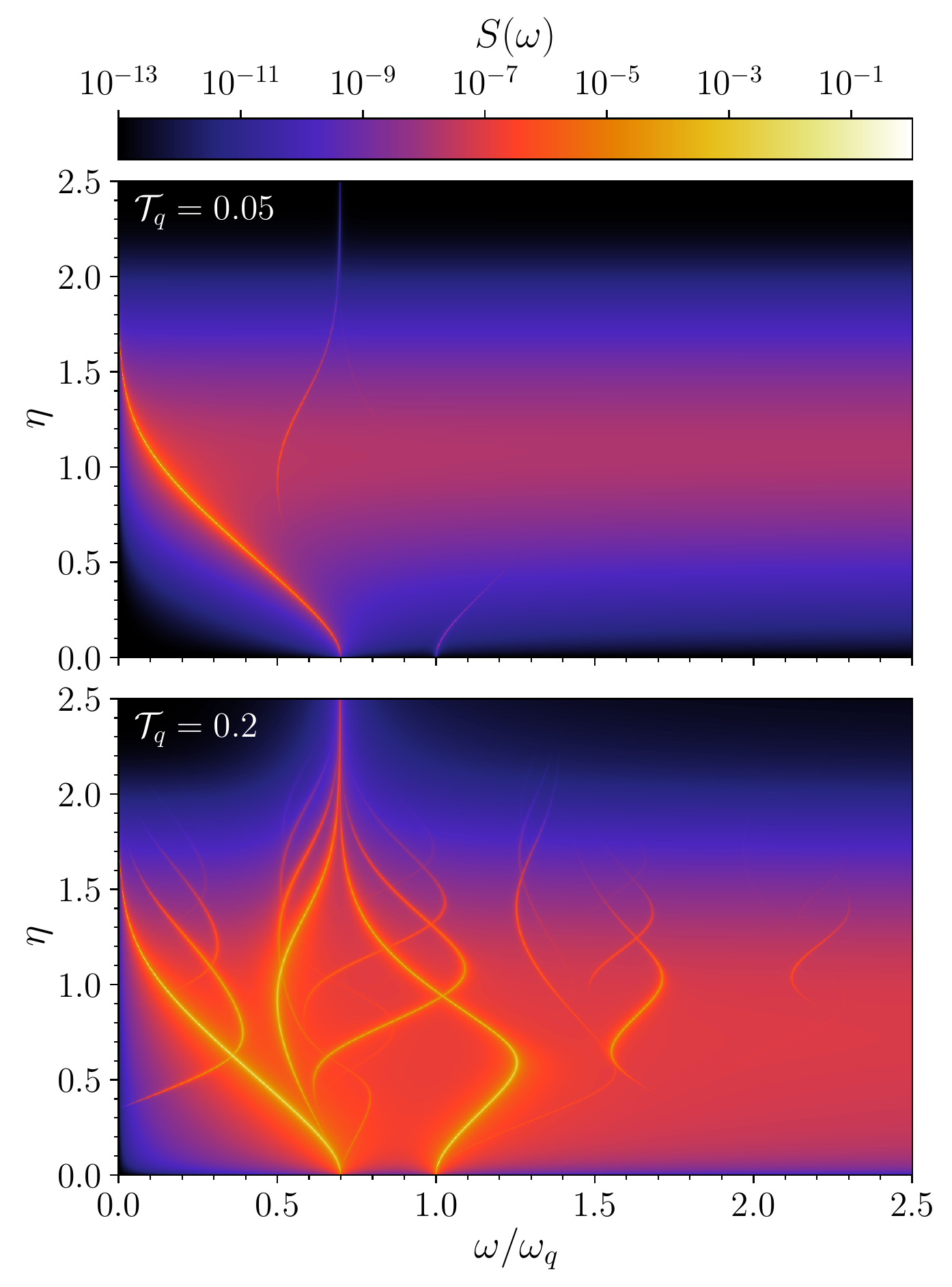}
    \caption{Cavity emission spectra $S_c(\omega)$ on logarithmic scale for values of $\eta$ reaching the USC and DSC regimes, obtained for two different effective qubit temperatures ${\cal T}_q = 0.05,\, 0.2$. We used $\Delta = -0.3$ ($\omega_c / \omega_q = 0.7$). The spectra have been normalized, so that the highest peak in each density plot is set at $1$. Increasing the temperature, additional lines originating from transitions involving higher energy levels appear, where the majority  of these correspond to transition energies shown in \figref{fig: energy_transitions_detuned_1}.}\label{fig: spectrum_strong_deep_detuned_1}
\end{figure}
The qubit emission rate is almost constant from the weak coupling regime to the onset of the USC, where it reaches values which are more than five orders of magnitude greater than the qubit emission rate at zero coupling $W_q^0$.
As in the zero detuning case, the strong enhancement of both the emission rates in the USC regime and at low effective temperature originates from the decrease of the transition frequency $\omega_{\tilde{1}_{-}, \tilde{0}}$, and as stated above, it is not present at higher effective temperatures.
Moreover, beyond the DSC regime onset ($\eta > 1$), both $\mathcal{W}_c$ and $\mathcal{W}_q$ decrease rapidly because of the light-matter and qubit-reservoir decoupling.

Figure \ref{fig: spectrum_weak_strong_detuned_1} shows the low-temperature ($\mathcal{T}_q = 5 \times 10^{-2}$) cavity and qubit emission spectra in the weak and strong coupling regimes ($10^{-5} \leq \eta \leq 3 \times 10^{-2}$), in analogy with the zero-detuning case in \figref{fig: spectrum_weak-strong}. 
As can be seen in \figref{fig: energy_transitions_detuned_1}, in the weak coupling regime $\omega_{\tilde{1}_{-},\tilde{0}} \approx \omega_c$ and $\omega_{\tilde{1}_{+},\tilde{0}} \approx \omega_q$. Moreover, at low values of the coupling strength and at low incoherent excitation rates (low temperature), the cavity cannot emit at the qubit frequency $\omega_q$, and the qubit cannot emit at $\omega_c$. Thus, \figref{fig: spectrum_weak_strong_detuned_1} shows only the $(\tilde{1}_{+}, \tilde{0})$ transition, while the $(\tilde{1}_{-}, \tilde{0})$ transition is absent.
\begin{figure}[ht]
    \centering
    \includegraphics[width = 0.48 \textwidth]{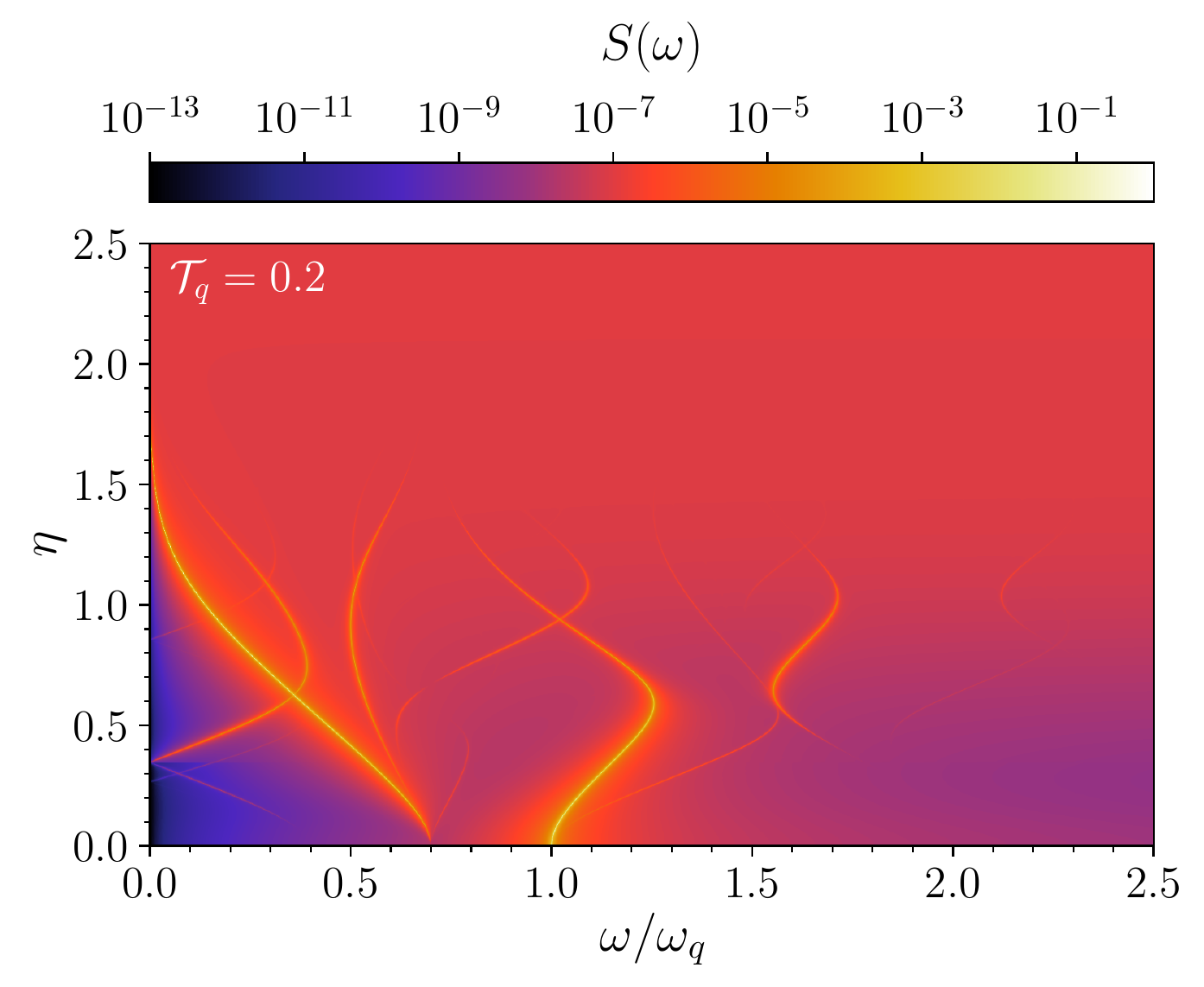}
    \caption{Logarithmic qubit emission spectra $S_q(\omega)$ for values of $\eta$ reaching the USC and DSC regimes, obtained at ${\cal T}_q = 0.2$. We used $\Delta = -0.3$. The spectra have been normalized, so that the highest peak is set at $1$. The visible lines  correspond to transition energies shown in \figref{fig: energy transitions}. The origin of the flat (red) signal background for $\eta > 1.5$ is explained in the text [see \eqref{backg}].} \label{fig: spectrum_strong_deep_qubit_detuned_1}
\end{figure}
Another difference between \figref{fig: spectrum_weak_strong_detuned_1} and \figref{fig: spectrum_weak-strong} is the intensity of the Purcell effect, which is much more effective in the zero-detuned case. Here the cavity starts to emit at $\eta \approx 10^{-2}$, a value which is three orders of magnitude greater than the zero detuning case.

Figure \ref{fig: spectrum_strong_deep_detuned_1} shows a logarithmic plot of the cavity emission spectra $ S_c( \omega)$ as a function of the normalized coupling strength $\eta$ ranging from the strong to the DSC regimes. As in the zero-detuning case, at very low effective temperature ($\mathcal{T}_q = 5 \times 10^{-2}$), the transition $(\tilde{1}_-, \tilde{0})$ is the brightest, while the atom-like transition $(\tilde{1}_+, \tilde{0})$ is visible only for $\eta \lesssim 0.4$. We also notice that the line at frequency $\omega_{\tilde{2}_-, \tilde{1}_-}$ becomes slightly visible only for $0.7 \lesssim \eta \lesssim 1.7$.

At $\mathcal{T}_q = 0.5$, additional energy levels get populated and additional spectral lines appear. As in the zero-detuning case, most of these transitions can be found in \figpanel{fig: energy_transitions_detuned_1}{b}. The line at frequency $\omega_{\tilde{2}_-, \tilde{1}_+}$ becomes sufficiently intense only for $\eta \gtrsim 0.35$, which is when the $(\tilde{1}_+, \tilde{0})$ frequency transition becomes greater than $(\tilde{2}_-, \tilde{0})$. We also observe that, at the higher effective temperature and at low coupling strengths (but with $\eta \gtrsim 0.05$), the cavity can also emit significantly at the qubit frequency $\omega_q$.
It is worth noting that at very high coupling strengths, all the spectral lines tend to be at frequencies which are multiple integer of the cavity frequency $\omega_c$.

Figure \ref{fig: spectrum_strong_deep_qubit_detuned_1} displays the logarithmic qubit emission spectra $S_q(\omega)$, which present features similar to those in the cavity emission spectrum but with a background emission above the onset of the deep USC regime.
\begin{figure}[ht!]
    \centering
    \includegraphics[width = 0.9 \linewidth]{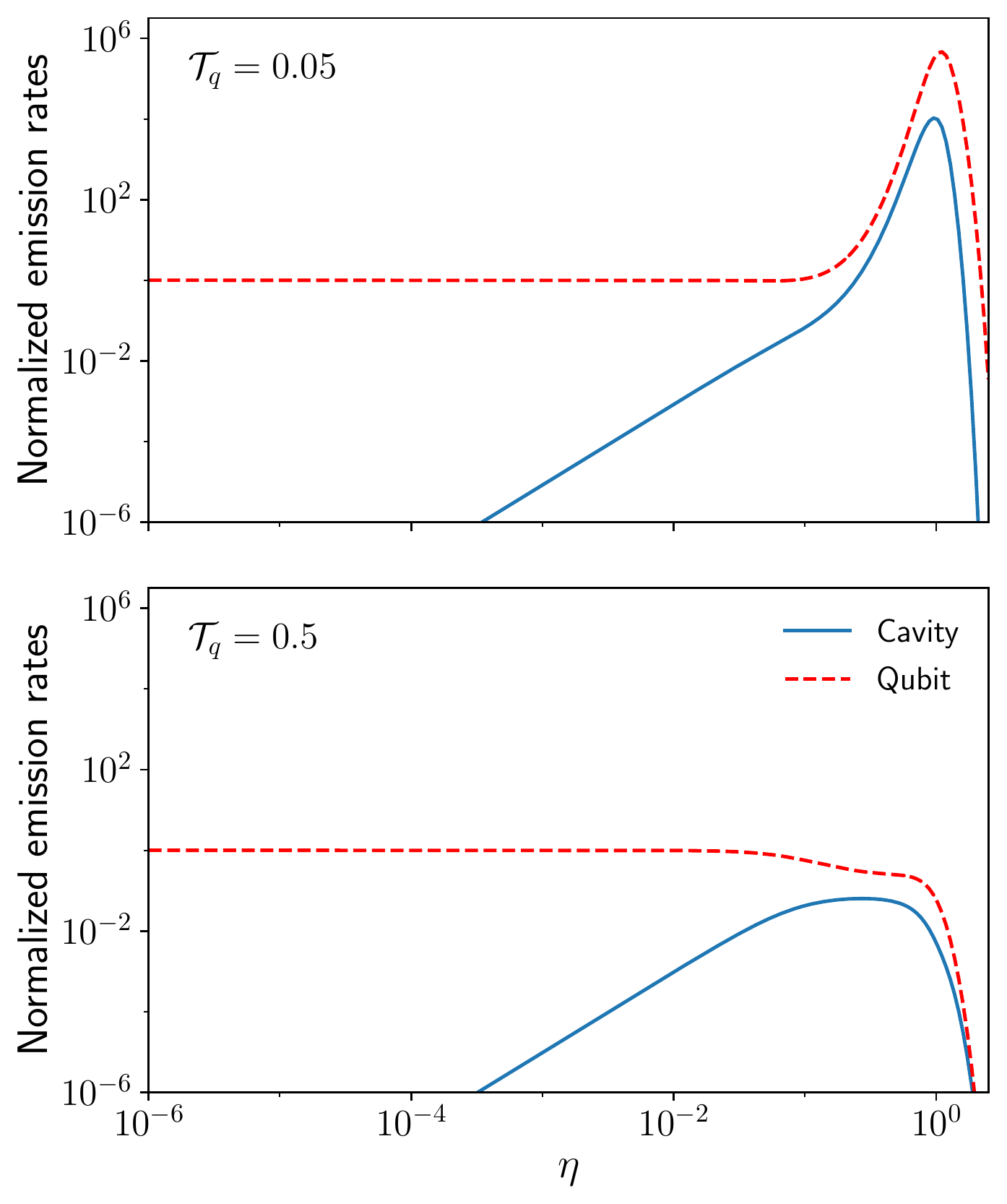}
    \caption{Cavity and qubit emission rates (normalized with respect to the qubit emission rate $W^0_q$ calculated for $\eta =0$) ${\cal W}_c =W_c/W^0_q$ (blue-continuous curve), and ${\cal W}_q = W_q/W^0_q$ (red-dashed) versus the light-matter normalized coupling strength $\eta$. We used $\Delta =0.3$, corresponding to $\omega_c/\omega_q = 1.3$. The upper panel has been obtained for ${\cal T}_q = 5 \times 10^{-2}$, the lower one for ${\cal T}_q = 5 \times 10^{-1}$.} \label{fig: ss_rates_detuned_2}
\end{figure}
\subsubsection{Cavity and qubit emission rates and spectra at positive detuning ($\Delta = 0.3$)}

The emission rates versus the normalized coupling strength $\eta$,  for the positive-detuning case, for both the cavity and the qubit are shown in \figref{fig: ss_rates_detuned_2} at both low and higher effective temperatures. For both the considered temperatures,  the cavity emission rate never exceeds the qubit one.
At $\mathcal{T}_q = 5 \times 10^{-2}$, the peak qubit emission rate exceeds the cavity peak by more than two orders of magnitude.
The increase of the cavity emission rate at increasing values of $\eta$ (Purcell effect) continues until the onset of the USC regime, and the plateau in the strong coupling regime, observed at zero-detuning, is here absent. 
At $\mathcal{T}_q = 5 \times 10^{-2}$, the qubit emission rate reaches in the USC regime values which are about six orders of magnitude larger than $W_q^0$. 

\begin{figure}[ht!]
    \centering
    \includegraphics[width = 0.9 \linewidth]{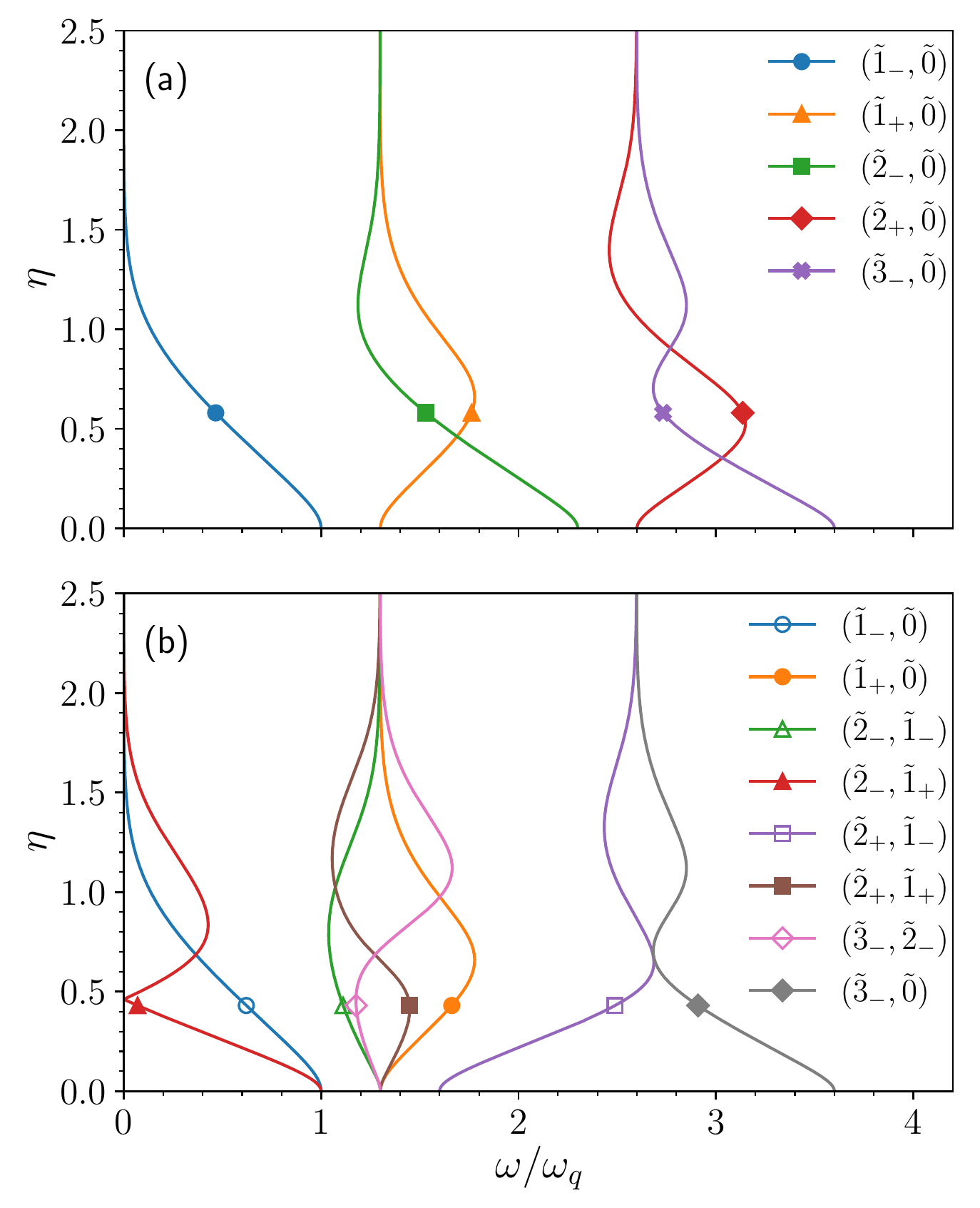}
    \caption{Normalized energy levels and transition energies versus $\eta$, for $\Delta = 0.3$ ($\omega_c / \omega_q = 1.3$). (a) lowest normalized energy levels (with the ground state energy as reference) $\omega_{{\tilde j}_\pm} - \omega_{\tilde 0}$ of the QRM. (b) Normalized parity-allowed transition energies $|\omega_{{\tilde j}_\pm} - \omega_{{\tilde k}_\pm}|$ for the lowest eigenstates of the QRM.} \label{fig: energy_transitions_detuned_2}
\end{figure}

\begin{figure}[ht!]
    \centering
    \includegraphics[width = 0.48 \textwidth]{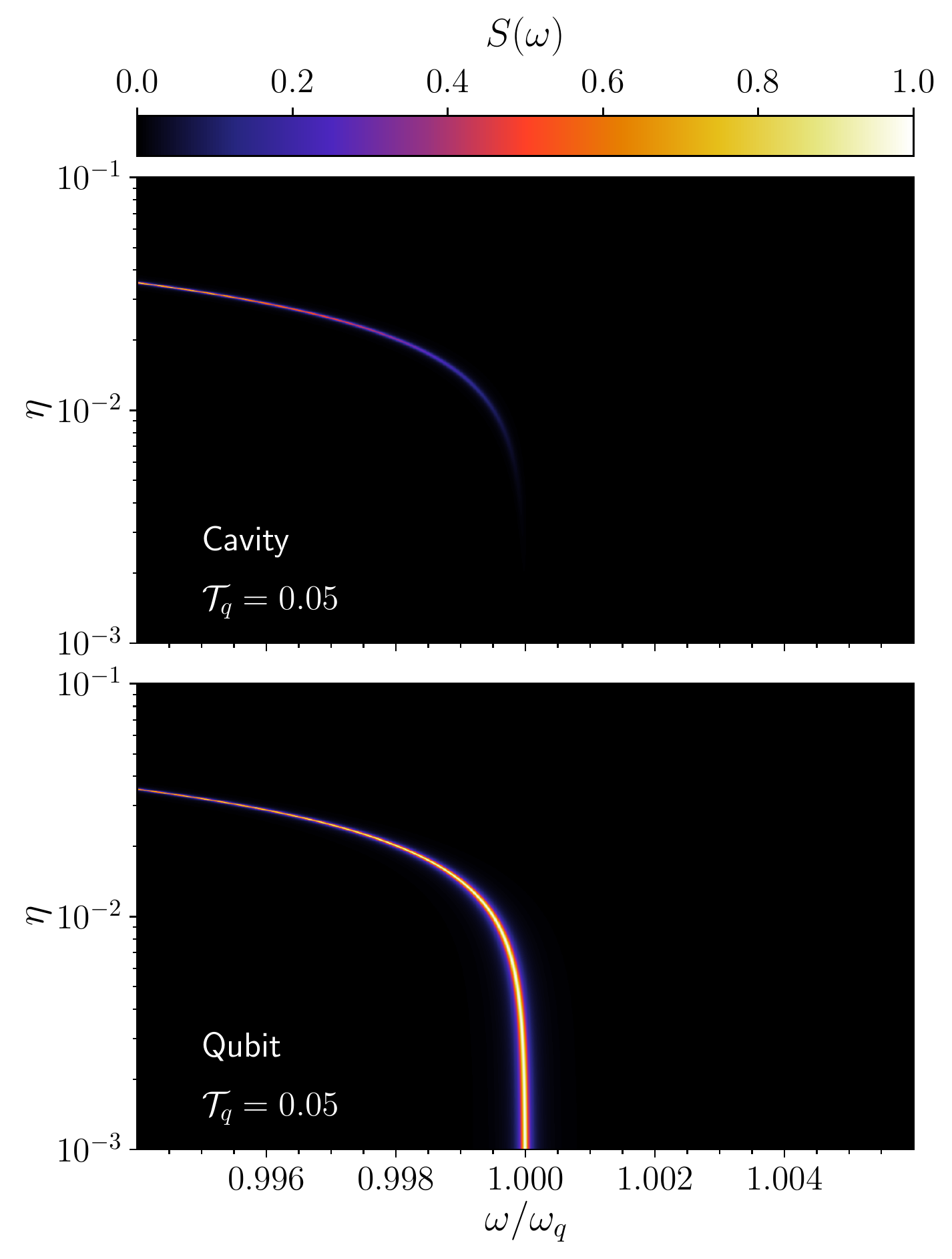}
    \caption{{Cavity $S_c(\omega)$ and qubit $S_q(\omega)$ emission spectra in the weak and strong coupling regime}, calculated for $10^{-3} < \eta < 10^{-1}$, and for $\Delta = 0.3$ ($\omega_c / \omega_q = 1.3$). The spectra have been obtained under weak incoherent excitation of the qubit. We used an effective qubit temperature ${\cal T}_q = 5 \times 10^{-2}$. The spectra have been normalized, so that the highest peak in each density plot is set at $1$.}
    \label{fig: spectrum_weak_strong_detuned_2}
\end{figure}

Looking at \figref{fig: energy_transitions_detuned_2}, we observe that, at weak coupling strengths the $(\tilde{1}_-, 0)$ transition almost coincides with the qubit transition frequency, while the $(\tilde{1}_+, 0)$ almost coincides with the resonance frequency of the cavity mode.
Figure \ref{fig: spectrum_weak_strong_detuned_2} shows the cavity emission spectrum, obtained at the effective temperature $\mathcal{T}_q = 5 \times 10^{-2}$, in the weak and strong coupling regimes.
In the frequency region displayed in \figref{fig: spectrum_weak_strong_detuned_2}, the only visible emission line corresponds to the transition $(\tilde{1}_-, \tilde{0})$.
As in the negative-detuning case (see  \figref{fig: spectrum_weak_strong_detuned_1}), in the weak coupling regime, the emission originates from the qubit-like transition $(\tilde{1}_-, 0)$, which however, now is the lowest energy transition.

\begin{figure}[ht]
    \centering
    \includegraphics[width = 0.48 \textwidth]{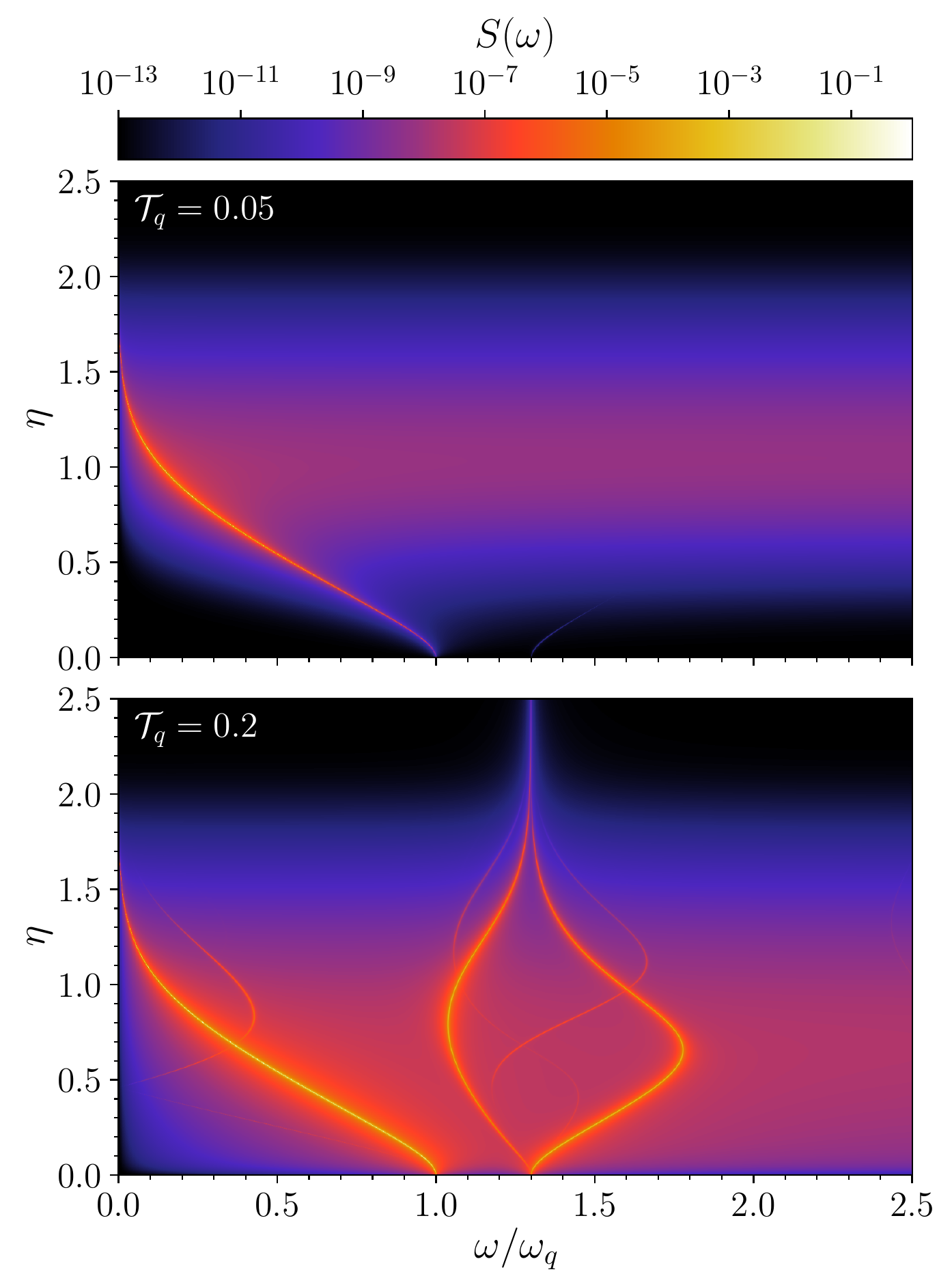}
    \caption{{ Cavity emission spectra $S_c(\omega)$ for values of $\eta$ reaching the USC and DSC regimes}, obtained for two different effective qubit temperatures ${\cal T}_q = 0.05,\, 0.2$. We used $\Delta = 0.3$ ($\omega_c / \omega_q = 1.3$). The spectra have been normalized, so that the highest peak in each density plot is set at $1$. Increasing the temperature, additional lines originating from transitions involving higher energy levels appear. Most of them correspond to transition energies shown in \figref{fig: energy_transitions_detuned_1}.}\label{fig: spectrum_strong_deep_detuned_2}
\end{figure}

\begin{figure}[ht]
    \centering
    \includegraphics[width = 0.48 \textwidth]{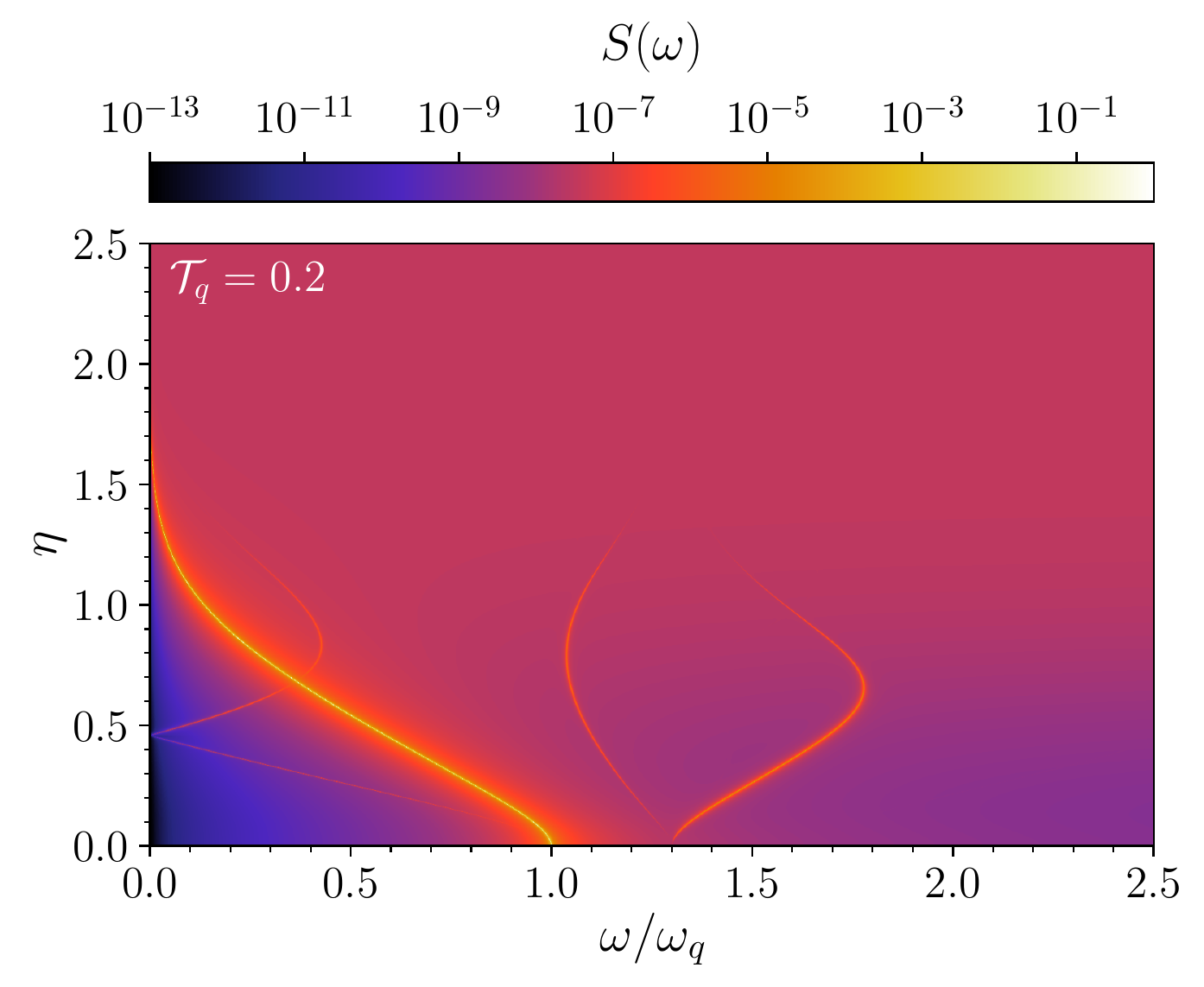}
    \caption{{Qubit emission spectra $S_q(\omega)$ for values of $\eta$ reaching the USC and DSC regimes}, obtained at ${\cal T}_q = 0.2$. We used $\Delta = 0.3$ ($\omega_c / \omega_q = 1.3$). The spectra have been normalized, so that the highest peak is set at $1$. The visible lines  correspond to transition energies shown in \figref{fig: energy transitions}. The origin of the flat (red) signal background for $\eta > 1.5$ is explained in the text [see \eqref{backg}]. }
    \label{fig: spectrum_strong_deep_qubit_detuned_2}
\end{figure}

Figure \ref{fig: spectrum_strong_deep_detuned_2} shows the logarithmic cavity emission spectra $S_c(\omega)$ as a function of the normalized coupling strength $\eta$, ranging from the strong to the DSC regimes. At very low effective temperatures, the emission originates almost only from the lowest energy transition $(\tilde{1}_-, 0)$, because the photon-like transition $(\tilde{1}_+, 0)$ is at higher energy and, owing to the detuning, is poorly hybridized with the qubit excited state, at least for moderate coupling strengths.
Increasing the effective temperature (${\cal T}_q = 0.2$) higher energy levels start to get populated, thus determining the appearance of several additional emission lines [see \figpanel{fig: energy_transitions_detuned_2}{b}].
The qubit  emission spectra versus $\eta$, at $\mathcal{T}_q = 0.2$ are shown in \figref{fig: spectrum_strong_deep_qubit_detuned_2}. The line corresponding to the transition $(\tilde{1}_+, 0)$ is the brightest at any coupling strength where emission lines are visibile. Also in this case the flat background qubit emission in the DSC regime dominates.
\section{\label{sec: Comparison with other models}Comparison with other models}
In this section, we present examples of calculations of cavity and qubit emission spectra, (using the same parameters adopted in the previous section) obtained using different models and/or dissipators for the master equation.
In particular, we present (i) some spectra obtained by using the JCM and the standard master equation for cavity QED, with the dissipators obtained by neglecting the light-matter interaction \cite{Beaudoin2011}; (ii)  we also present spectra obtained with the same model used in Sect. \ref{sec: Results}, using a master equation where the dissipators have been obtained taking into account the light-matter interaction, but applying the post-trace RWA \cite{Beaudoin2011}.
\begin{figure}[ht]
    \centering
    \includegraphics[width = 0.48 \textwidth]{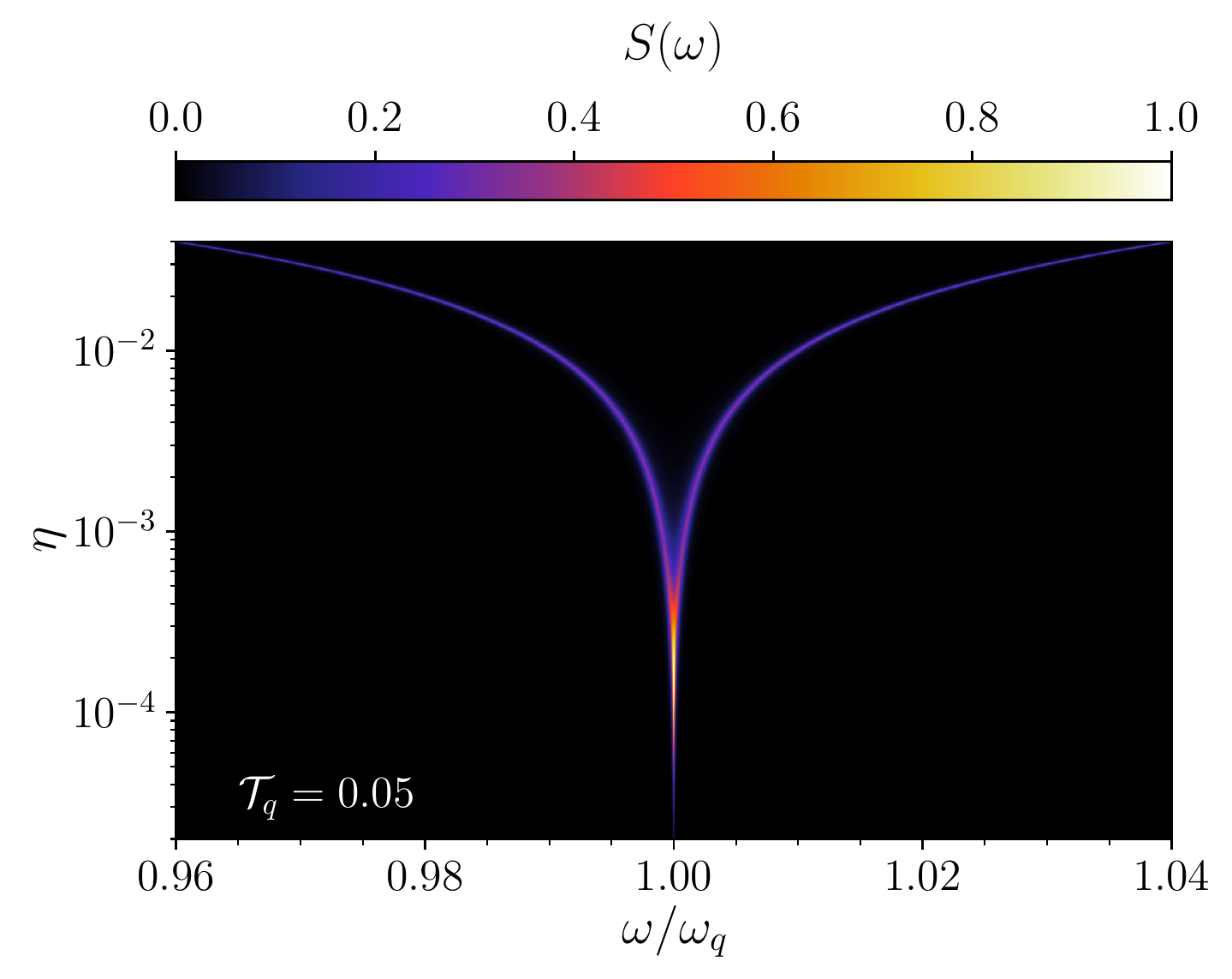}
    \caption{Cavity emission spectra $S_c(\omega)$ in the weak and strong coupling regime for the JCM, calculated for $2 \times 10^{-5} < \eta < 4 \times 10^{-2}$, and for $\Delta = 0$. The spectra have been obtained under weak incoherent excitation of the qubit. We used an effective qubit temperature ${\cal T}_q = 5 \times 10^{-2}$. The spectra have been normalized, so that the highest peak in each density plot is set at $1$.} \label{fig: spectrum_weak_strong_JC}
\end{figure}
\subsection{\label{subsec: Standard master equation using James-Cummings Hamiltonian}Jaynes-Cummings model}
The JCM is the simplest model describing a two level system interacting with a quantized single-mode of an electromagnetic resonator. It is obtained applying the RWA to the QRM in the dipole gauge. Therefore, it is expected to be valid only for coupling strengths below the USC regime.
The JC Hamiltonian is
\be
\label{eq: JC Hamiltonian}
\hat{\mathcal{H}}_{JC} = \omega_c \adop \aop + \frac{\omega_q}{2} \sz + \frac{\eta \omega_c}{2} \left( \aop \sp + \adop \sm \right)\, .
\ee

In addition, in order to describe the interaction of the qubit-cavity system with the environment, we used the standard quantum optical master equation
\be
\dot{\hat{\rho}} = -i \comm{\hat{\mathcal{H}}_{JC}}{\hat{\rho}} + \mathcal{L}^{\text c}_{\text{bare}} \hat{\rho}
+ \mathcal{L}^{\text q}_{\text{bare}} \hat{\rho}\, ,
\ee
where $\mathcal{L}^{\text c}_{\text{bare}}$ and $\mathcal{L}^{\text q}_{\text{bare}}$
are the standard dissipators for the cavity and the qubit respectively:
\begin{figure}[ht]
    \centering
    \includegraphics[width = 0.48 \textwidth]{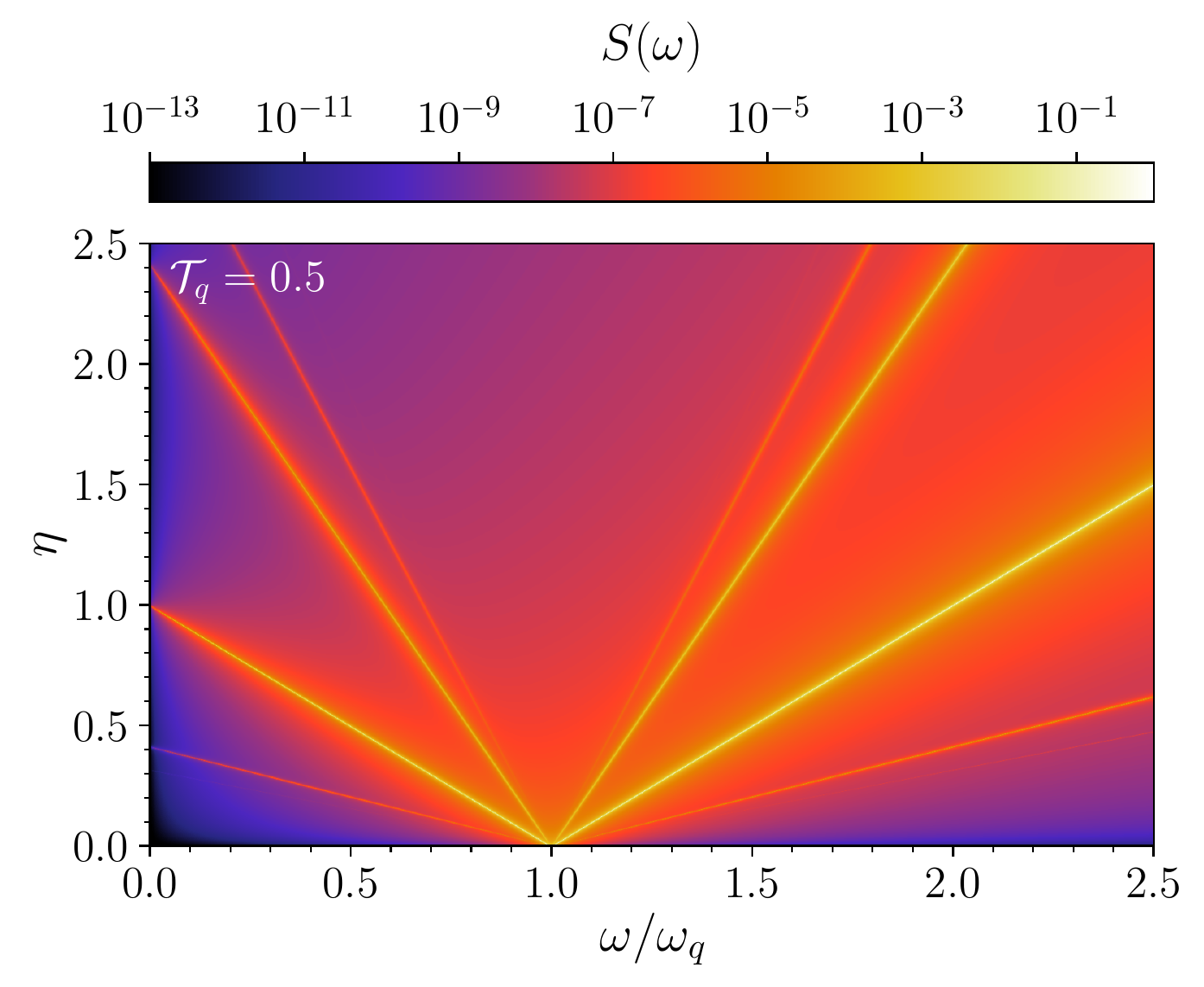}
    \caption{Cavity emission spectra $S_c(\omega)$ for values of $\eta$ reaching the USC and DSC regimes with the JCM, obtained with an effective qubit temperature ${\cal T}_q = 0.5$. We used $\Delta = 0$. The spectra have been normalized, so that the highest peak in each density plot is set at $1$. As predicted, the model drastically fails above the USC regime.}
    \label{fig: spectrum_strong_deep_JC}
\end{figure}
\bea
\label{eq: standard Lindbaldian superoperator}
\mathcal{L}^{\text c}_{\text{bare}} \hat{\rho} &=& \kappa \left[ 1 + n_c(\mathcal{T}_c) \right] \mathcal{D} \left[ \aop \vphantom{\adop} \right] \hat{\rho} \nonumber \\
&+& \kappa n_c(\mathcal{T}_c) \mathcal{D} \left[ \adop \vphantom{\adop} \right] \hat{\rho} \nonumber \\
\mathcal{L}^{\text q}_{\text{bare}} \hat{\rho} &=& \gamma \left[ 1 + n_q(\mathcal{T}_q) \right] \mathcal{D} \left[ \sm \vphantom{\adop} \right] \hat{\rho} \nonumber \\
&+& \gamma n_q(\mathcal{T}_q) \mathcal{D} \left[ \sp \vphantom{\adop} \right] \hat{\rho}\, .
\eea

The term  $n_{c (q)}(\mathcal{T}_{c (q)}) = [\exp(1 / \mathcal{T}_{c(q)}) - 1]^{-1}$ is the thermal population, and $\mathcal{D} [ \hat O ]$ indicates the generic dissipator
\be
\label{eq: generic dissipator}
\mathcal{D} \left[ \hat{O} \right] \hat{\rho} = \frac{1}{2} \left(2 \hat{O} \hat{\rho} \hat{O}^\dagger - \hat{\rho} \hat{O}^\dagger \hat{O} - \hat{O}^\dagger \hat{O} \hat{\rho} \right)\, .
\ee
In this case, the cavity and qubit emission rates are simply proportional to $W_c(t) = \langle \hat a^\dag \hat a \rangle_t$ and
$W_q(t) = \langle \sp \sm  \rangle_t$, respectively. Analogously, the steady-state cavity and qubit emission spectra can be defined as
\bea
\tilde S_c(\omega) &=& {\rm Re} \int_0^\infty d \tau  e^{-i \omega \tau}\langle \hat a^\dag(t+ \tau) \hat {a}(t) \rangle_{ss} \nonumber \\
\tilde S_q(\omega) &=& {\rm Re} \int_0^\infty d \tau  e^{-i \omega \tau}\langle \sp (t+ \tau) \sm (t) \rangle_{ss}\, .
\label{spectra}\eea

Figure \ref{fig: spectrum_weak_strong_JC} describes the normalized emission spectra of the cavity in the weak-strong coupling range. As mentioned above, the JCM is a good approximation in this range of coupling strengths. The only difference with respect to the results obtained using the full QRM (see \figref{fig: spectrum_weak-strong}) is the lack of any intensity difference between the split lines originating from the transitions $({1}_{-}, 0)$ and $({1}_{+}, 0)$. Actually, the reason is not directly due to the use of the JCM, but is a consequence of using the standard master equation which does not include  bath populations calculated at the system transition frequencies.
Of course such a model cannot provide reliable results beyond the strong coupling regime (see \figref{fig: spectrum_strong_deep_JC}).

\subsection{\label{subsec: Blais}Dressed master equation with post-trace RWA}
  The standard quantum-optics master equation has several weaknesses, and one of the most relevant is that  the interaction between the subsystems is not considered when deriving the dissipators \cite{carmichael1973master}. 
One of the main drawbacks is that, owing to the presence of counter-rotating terms in the Hamiltonian describing the interaction between the light and matter components of the system, unphysical excitations are generated into the system even by a zero-temperature reservoir.
For this reason a dressed master equation was developed \cite{Beaudoin2011}. This model takes into account that transitions in the hybrid system occur between dressed eigenstates, and not between the
eigenstates of the free Hamiltonians of the components.

The dressed-state master equation is
\be
\dot{\hat{\rho}} = -i \comm{\hat{\mathcal{H}}_R}{\hat{\rho}} + \mathcal{L}_{\text{dressed}} \hat{\rho}\, ,
\ee
where $\mathcal{L}_{\text{dressed}}$ is of the form 
\begin{figure}[ht]
    \centering
    \includegraphics[width = 0.48 \textwidth]{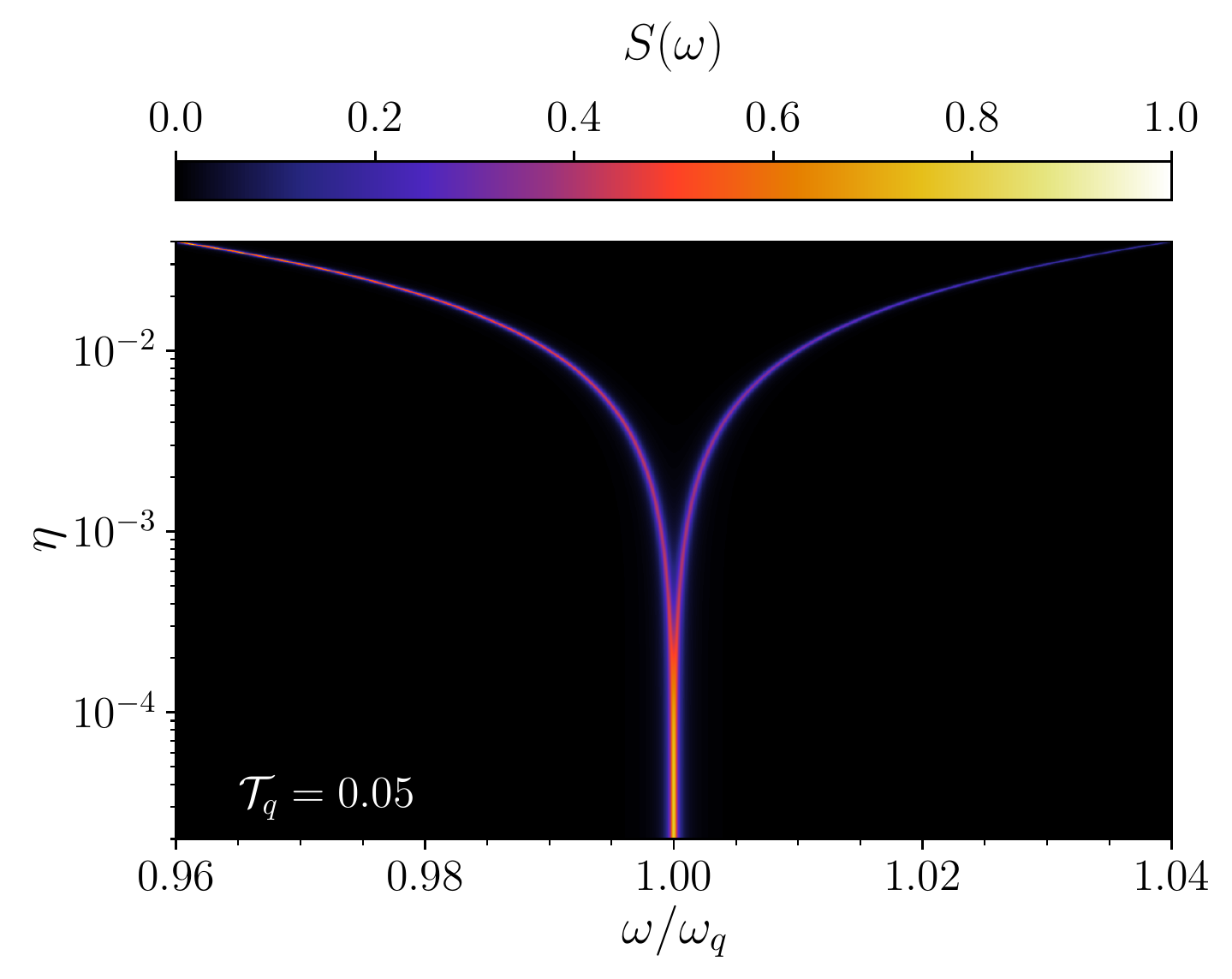}
    \caption{Cavity emission spectra $S_c(\omega)$ in the weak and strong coupling regime using the dressed master equation, calculated for $2 \times 10^{-5} < \eta < 4 \times 10^{-2}$, and for $\Delta = 0$. The spectra have been obtained under weak incoherent excitation of the qubit. We used an effective qubit temperature ${\cal T}_q = 5 \times 10^{-2}$. The spectra have been normalized, so that the highest peak in each density plot is set at $1$.} \label{fig: spectrum_weak_strong_blais}
\end{figure}
\begin{figure}[ht]
    \centering
    \includegraphics[width = 0.48 \textwidth]{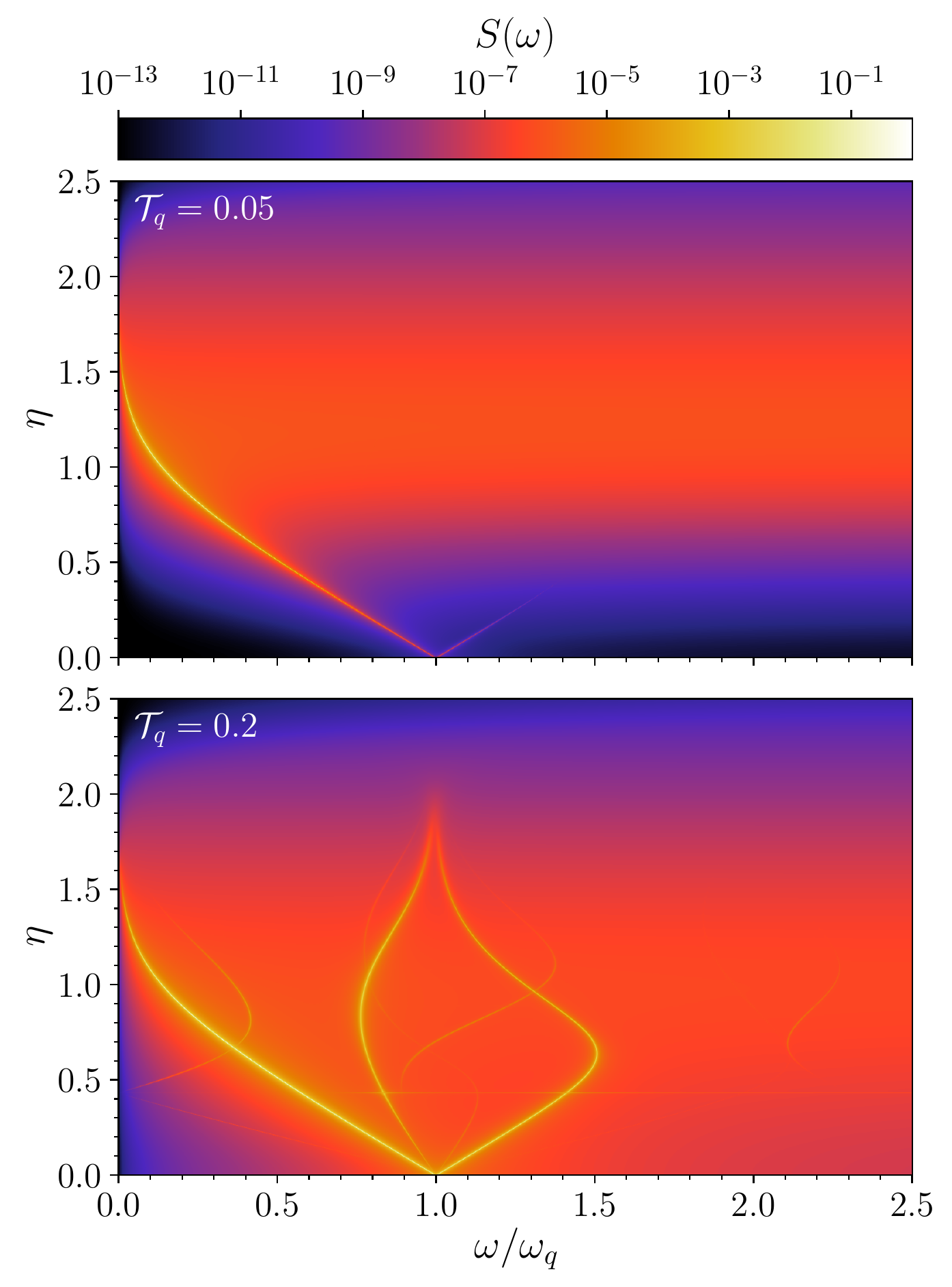}
    \caption{Cavity emission spectra $S_c(\omega)$ for values of $\eta$ reaching the USC and DSC regimes using the dressed master equation, obtained with an effective qubit temperature ${\cal T}_q = 0.5$. We used $\Delta = 0$. The spectra have been normalized, so that the highest peak in each density plot is set at $1$. In the DSC regime (when the system became harmonic) this model fails.}
    \label{fig: spectrum_strong_deep_blais}
\end{figure}
\bea
\label{eq: dressed Lindbaldian superoperator}
\mathcal{L}_{\text{dressed}} \hat{\rho} &=& \sum_{i=c,q} \sum_{k > j} \left\{ \Gamma_{kj}^{i} \left[ 1 + n_{kj}(\mathcal{T}_i) \right] \mathcal{D} \left[ \dyad{j}{k} \right] \hat{\rho} \right. \nonumber \\
&+& \left. \Gamma_{kj}^{i} n_{kj}(\mathcal{T}_i) \mathcal{D} \left[ \dyad{k}{j} \right] \hat{\rho} \right\}.
\eea
The terms $n_{kj}(\mathcal{T}_i)$ are the thermal reservoir populations [see \eqref{NT}], calculated at the system transition frequencies $\omega_{kj}$,  and  $\Gamma_{kj}^{c(q)}$ are the cavity and qubit decay [see \eqref{Gc1} and \eqref{Gq1}].

As mentioned above, this dissipation model is valid only with anharmonic systems and, as can be noticed in \figref{fig: spectrum_weak_strong_blais}, it returns incorrect results when the coupling is very weak and the system displays a quasi harmonic spectrum---that is a non zero cavity photon emission at negligible couplings [compare with  \figpanel{fig: spectrum_weak-strong}{a}].

Figure \ref{fig: spectrum_strong_deep_blais} shows the normalized emission spectra of the cavity, using the dressed master equation in the USC-DSC range. 
Also at very large values of the coupling strength, when the energy spectrum of the QRM tends towards harmonicity,  the obtained emission spectra shown in \figref{fig: spectrum_strong_deep_blais}, differ significantly from the corresponding ones in \figref{fig: spectrum_strong-deep}. The main difference is the background emission which covers almost all the USC and DSC region. Moreover, the emission lines are not well defined in the DSC region.

\section{\label{sec: Conclusion}Conclusions}
We have investigated how the QRM emits light under incoherent excitation of the two-level atom, considering coupling strengths ranging from the weak coupling to the USC and DSC regimes. We analyzed both the cavity and the qubit emission, for both the resonant and detuned cases, considering different effective qubit-temperatures.
In particular, by using a GME approach, we were able to calculate numerically cavity and qubit emission rates and spectra versus the normalized light-matter coupling strength and for different incoherent qubit excitation strengths (effective temperature).
Following the work in Ref. \cite{salmon2021gauge},
the obtained results are gauge independent. 
The theoretical framework 
allows us to investigate the light-matter decoupling and the fate of the Purcell effect in the QRM when the normalized coupling strength $\eta$ is significantly larger than one. In this case, we found that the cavity and qubit emission rates are affected both by  light-matter decoupling and  qubit-reservoir decoupling.

Reaching the USC and DSC regimes with individual quantum emitters, whose interaction with light is implemented via the minimal coupling replacement, is currently very difficult,
though progress is being made, especially with 
plasmonic cavity systems~\cite{Mueller2020}.
However, this theoretical framework can be easily generalized to include $N$ qubits (Dicke model) \cite{GarzianoPRA2020}.
Moreover, coupling strengths ranging from the weak to the DSC regime can be achieved with individual qubits using superconducting circuits (see, e.g., \cite{Niemczyk2010, Yoshihara2017}). In this case, however, the light-matter interaction is not described by the minimal coupling replacement and specific calculations for these systems would have to be carried out, where the general approach described here and these results can provide a precise guide.

\section{\label{sec: Acknowledgments}Acknowledgments}
The first two authors (A.M. and V.M) equally contributed to this work.
F.N. is supported in part by: Nippon Telegraph and Telephone Corporation (NTT) Research, the Japan Science and Technology Agency (JST) [via the Quantum Leap Flagship Program (Q-LEAP), the Moonshot R$\&$D Grant Number JPMJMS2061, and the Centers of Research Excellence in Science and Technology (CREST) Grant No. JPMJCR1676], the Japan Society for the Promotion of Science (JSPS) [via the Grants-in-Aid for Scientific Research (KAKENHI) Grant No. JP20H00134 and the JSPS–RFBR Grant No. JPJSBP120194828], the Army Research Office (ARO) (Grant No. W911NF-18-1-0358), the Asian Office of Aerospace Research and Development (AOARD) (via Grant No. FA2386-20-1-4069), and the Foundational Questions Institute Fund (FQXi) via Grant No. FQXi-IAF19-06. 
S.S. acknowledges the Army Research Office (ARO) (Grant No. W911NF-19-1-0065).
S.H. thanks
 the Natural Sciences and Engineering Research Council of Canada, the Canadian Foundation for Innovation, the National Research Council of Canada, and Queen's University.

\appendix

\section{Cavity- and qubit-bath interactions from the gauge principle }\label{Gauge_issues_cavity_qubit-bath_interaction}
A simple and widely adopted  way to model the interaction of a quantum system with its environment,  
consists in writing the interaction Hamiltonian as a quadratic form involving the product of system- and bath-degrees of freedom
\begin{equation}\label{App:Hamiltionian_SB}
   \hat H_{\rm I} =
    \sum_{k}
    \lambda_{k}\hat S  \hat B_k\,,
\end{equation}
where the coefficients $\lambda_{k}$ represent the coupling strengths (assumed real, one for any bath-degree of freedom), while $\hat S$ and $\hat B_k$ are Hermitian system- and bath-operators.  
The specific values of the coupling strengths and the choice of the system and bath operators in \eqref{App:Hamiltionian_SB} is model dependent. In the next subsections, we present arguments 
to guide and to make these choices consistent  (taking care of gauge issues).  Specifically, we present a consistent derivation for the QRM, describing the interaction of the cavity mode and the qubit with their respective baths.

The approach followed here is close to that recently developed in Ref.~\cite{salmon2021gauge}, and gives rise to analogous results. The only difference is that here we introduce the interaction of the photonic and matter components with their reservoirs by invoking the gauge principle. This approach provides a rather fundamental model of systems-baths interactions and  connects more with quantum field theory.

\subsection{Cavity-bath interaction in the Coulomb and  dipole gauges}

Here, we introduce the cavity-bath coupling
by regarding the environment as collective bosonic excitations of  matter, so that we can introduce such interaction by invoking the fundamental gauge principle, which determines the form of light-matter coupling.

\subsubsection{ Cavity-bath interaction in the Coulomb gauge}\label{app:cavity_bath_interaction_coulomb}

As a consequence of the gauge principle (minimal coupling replacement), in the Coulomb gauge, the only degrees of freedom of the electromagnetic field interacting with matter are the field coordinates. For a single mode resonator, there is a single coordinate $\hat x$, which can be expressed as  $\hat x=\hat a + \hat a^\dag$. 
This fact not only leads to the useful relation in \eqref{sec:matrix_elemet_coulomb}, but it also allows to obtain a  Thomas-Reiche-Kuhn (TRK) sum rule for the electromagnetic field \cite{savastaNano2021}.

Usually, input-output frameworks, used to model the interaction of cavity modes with the external modes, adopt as bath operators a continuum of bosonic electromagnetic modes, rather than a collection of bosonic matter excitations. However, 
the field-bath coupling, ultimately, originates from the interaction of the cavity electromagnetic field with matter (e.g., cavity mirrors). Due to this fact,  even if such matter degrees of freedom can be adiabatically eliminated,
it is reasonable to require that the general relationship \eqref{sec:matrix_elemet_coulomb} and the TRK sum rule remain valid. To ensure this, it is sufficient to start from the free bosonic Hamiltonian for the bath,  $\hat H_b=\sum_k\omega_k\hat b_k^{\dagger} \hat b_k $, and to apply the {\em generalized minimal coupling replacement}, corresponding to transforming each matter operator of the bath by a suitable unitary transformation
(see Ref.~\cite{GarzianoPRA2020}):
\be
\hat b_k \to \hat {\cal U}_{cb}\, \hat b_k\, \hat {\cal U}_{cb}^\dag= \hat b_k - i \eta_k^c (\hat a + \hat a^\dag)\, ,
\ee
where
\be\label{App:Gauge_trasformation}
\hat {\cal U}_{\rm cb} = \exp \left[ i(\hat a + \hat a^\dag)  \sum_k\eta^{\rm c}_k (\hat b_k + \hat b^\dag_k) \right]\, .
\ee
Assuming a coupling constant ({effective charge} $q$) independent on the matter mode $k$, 
$\hat {\cal U}_{\rm cb}$ can be written as
\be\label{T2}
\hat {\cal U}_{\rm cb} = \exp \left[iq \hat x\, \hat Q_{\rm b}
\right]\, ,
\ee
where the coordinate of the reservoir field reads
\be\label{QM}
\hat Q_{\rm b} =  \sum_k \frac{\alpha}{\sqrt{\omega_k}}  (\hat b_k + \hat b^\dag_k)\, .
\ee
Here $\alpha$ is a constant. The resulting  adimensional coupling in \eqref{App:Gauge_trasformation} can be expressed as
\be\label{eta}
\eta_k^{\rm c} = q \alpha / \sqrt{\omega_k}\, .
\ee
Of course, depending on the specific model, the effective charge can be frequency dependent, implying a different frequency dependence of $\eta^c_k$  than that in \eqref{eta}.
By neglecting the diamagnetic term [of the second order ${\eta^{\rm c}_k}^2(\omega_k)$], we obtain the interaction Hamiltonian for the cavity field and the reservoir 
\be\label{App:Cavity_bath_interaction}
\hat H_{\rm cb} = i  (\hat a + \hat a^\dag) \sum_k  \omega_k \eta^{\rm c}_k (\hat b_k- \hat b^\dag_k)\, .
\ee
A comparison between \eqref{App:Hamiltionian_SB} and \eqref{App:Cavity_bath_interaction} clearly shows that, $\hat S= (\hat a + \hat a^\dag)$ and $\hat B_k= i \omega_k  (\hat b_k- \hat b^\dag_k)$.

Equation (\ref{App:Cavity_bath_interaction}) represents a consistent starting point to study the interaction of the cavity mode with its bath. In order to derive the GME (see \appref{app: generalized master equations}), it is useful to expand the cavity-field coordinate in the dressed basis 
\be\label{VC02}
\hat H_{\rm cb} = i \sum_{l,m} x_{lm} \hat P_{lm} \sum_k  \omega_k \eta_k^c (\hat b_k- \hat b^\dag_k)\, ,
\ee
where $\hat P_{lm} = | l \rangle \langle m|$  are the transition operators and  $x_{lm} = \langle l |( \hat a + \hat a^\dag )| m \rangle$ are the matrix elements for the cavity-field coordinate, being $|l \rangle$ the energy eigenstates of the Hamiltonian in the Coulomb gauge [\eqref{eq: coulomb_gauge_rabi_hamiltonian} in \secref{sec: Theoretical_Model_Coulomb_gauge}]. 
In terms of cavity-photon operators with positive and negative frequencies \cite{Ridolfo2012}, \eqref{VC02} can be written as
\be\label{VC2}
\hat H_{\rm cb} = i \sum_{l>m} \bigg (   x_{lm} \hat P_{lm} +  x_{ml}^*\hat P_{ml} \bigg ) \sum_k  \omega_k \eta_k^c (\hat b_k- \hat b^\dag_k)\, .
\ee
Note that, $\sum_{l>m} x_{lm} \hat P_{lm}= ( \sum_{l<m} x_{ml}^*\hat P_{ml})^{\dagger}$.

Since the system-bath  interaction  is assumed to be weak, it is reasonable to apply to  \eqref{VC2} the rotating wave approximation:
\be\label{VC2n}
\hat H_{\rm cb} = -i \sum_{l >m} x_{ml}^* \hat P_{ml}  \sum_k  \omega_k \eta_k^c  \hat b^\dag_k + {\rm H.c.}\, .
\ee

\subsubsection{Cavity-bath interaction in the dipole gauge}
The cavity-bath interaction Hamiltonian in the dipole gauge 
could be obtained by simply performing the gauge transformation of the cavity operators in \eqref{App:Cavity_bath_interaction}:
$\hat a \to \hat {\cal R} \hat a \hat {\cal R}^\dag = \hat a + i \eta^c \hat \sigma_x=\hat a'$, where 
\be\label{App:Gauge_trasformation_reduced}
\hat {\cal R} = \exp \left [ -i\eta (\hat a + \hat a^\dag) \hat \sigma_x
\right ]\, .
\ee
Therefore, $\hat H'_{\rm cb}$ in the dipole gauge is equivalent to  $\hat H_{\rm cb}$ in the Coulomb gauge, because $\hat x' = \hat {\cal R} \hat x \hat {\cal R}^\dag$.
In doing so, we are neglecting the action of the gauge transformation on the reservoir operators $\hat b_k $ ($\hat b_k^{\dagger}$), which is possible because,  as shown in Ref.~\cite{DiStefano2019a}, gauge transformations affect significantly ladder operators only when the coupling is rather strong. 
Nevertheless, it remains interesting and instructive to derive directly the cavity-bath reservoir in the dipole gauge. 

In general, for a light-matter system whose light component (in the absence of interaction)  is described by the Hamiltonian $\hat H_{\rm ph}$, and matter component by $\hat H_m$, the total Hamiltonian (in the dipole approximation) in the presence of interaction can be obtained in the Coulomb gauge by applying a suitable unitary transformation $\hat U$ to the matter Hamiltonian only:
\be
\hat H = \hat U \hat H_m \hat U^\dag + \hat H_{\rm ph}\, .
\ee
The dipole gauge Hamiltonian  is obtained from the above as
\be
\hat H' = \hat U^\dag \hat H_C \hat U = \hat H_m + \hat U^\dag \hat H_{\rm ph} \hat U\, ,
\ee
which corresponds to applying a unitary transformation to the bare photonic Hamiltonian only: a sort of generalized minimal coupling replacement for the photonic Hamiltonian.
Of course, the dipole-gauge Hamiltonian can also be written  using the unitary operator $\hat R = \hat U^\dag$:
\be
\hat H' = \hat H_m + \hat R \hat H_{\rm ph} \hat R^\dag\, .
\ee

In the present case,  the generalized minimal coupling replacement for a single-mode photonic system interacting with a two-level atom and with a bosonic matter field, can be implemented by the following unitary operator
\be\label{App:Gauge_trasformationR}
\hat {\cal R}_{\rm cqb} = \exp \left\{ -i(\hat a + \hat a^\dag) [ \eta \sigma_x +  \sum_k\eta^{\rm c}_k (\hat b_k + \hat b^\dag_k) ]\right\}\, .
\ee
The dipole-gauge Hamiltonian for the system constituted by a cavity-mode interacting with a qubit and with a bosonic matter reservoir can be directly obtained  applying the generalized gauge transformation  [see \eqref{App:Gauge_trasformationR}] to the photon operators in the bare cavity Hamiltonian $\hat H_{\rm ph} = \omega_c \hat a^\dag \hat a$:
\be
\hat a \to \hat {\cal R}_{cqb} \hat a \hat {\cal R}_{cqb}^\dag = \hat a + i \eta \hat \sigma_x + i \sum_k \eta^{\rm c}_k (\hat b_k + \hat b^\dag_k)\, .
\ee
The resulting cavity-bath interaction Hamiltonian in the dipole gauge becomes
\bea
\hat H'_{\rm cb} &=& -i \omega_c (\hat a - \hat a^\dag + 2i \eta \hat \sigma_x) \sum_k \eta^{\rm c}_k (\hat b_k+ \hat b^\dag_k)\nonumber \\
&+& \omega_c \left| \sum_k \eta^{\rm c}_k (\hat b_k+ \hat b^\dag_k)  \right|^2\, ,
\label{VD}\eea
which includes an effective interaction of the bath with the qubit, and a bath self-interaction term.
The result in \eqref{VD} differs from what has been obtained above adopting a less rigorous approach ($\hat H'_{\rm cb} = \hat H_{\rm cb}$). However, in \appref{app: generalized master equations} we show that the two approaches, after the Markov and Born approximations, give rise to the same dissipators in the master equation.
Neglecting the bath self-interaction term [of second order $\eta_k^2(\omega_k)$], and using the dipole-gauge photon operators in \eqref{dipolo_photon_operator} (see \secref{sec: Theoretical_Model_dipole_gauge}), the cavity-bath interaction Hamiltonian in \eqref{VD} can be written as
\be\label{VD1}
\hat H'_{\rm cb} = -i \omega_c (\hat a' - \hat a'^\dag) \sum_k \eta^{\rm c}_k (\hat b_k+ \hat b^\dag_k)\, .
\ee
A comparison between \eqref{App:Hamiltionian_SB} and \eqref{VD1} clearly shows that, $\hat S= -i \omega_c (\hat a' - \hat a'^{\dag})$ and $\hat B_k= \hat b_k + \hat b_k^\dag$.

By expanding in \eqref{VD1} the cavity operators  in the dressed basis, we obtain
\be\label{VD2}
\hat H'_{\rm cb} = -i \omega_c \sum_{m,l} p'_{lm} \hat P'_{lm} \sum_k \eta^{\rm c}_k (\hat b_k+ \hat b^\dag_k)\, ,
\ee
where $p'_{lm} =  \langle l'|( \hat a' - \hat a'^\dag) |m' \rangle$, and $\hat P'_{lm} = |l' \rangle \langle m'|$, being $|l' \rangle$ the energy eigenstates of the Rabi  Hamiltonian in dipole gauge [\eqref{eq: dipole_gauge_rabi_hamiltonian} in \secref{sec: Theoretical_Model_dipole_gauge}].
Looking at \eqref{sec:matrix_elemet_dipole} in the main text, 
the above matrix element can also be written as $p'_{lm} = \omega_{lm} x'_{lm}/\omega_c$, with $ x'_{lm}=\langle l'|( \hat a' + \hat a'^\dag) |m' \rangle$. In terms of cavity-photon operators with positive and negative frequencies, we obtain
\be\label{VD3}
\hat H'_{\rm cb}= -i \sum_{l>m} \omega_{lm}\bigg ( x'_{lm} { \hat P'}_{lm} -  x'^*_{ml} {\hat P'}_{ml} \bigg ) \sum_k  \eta^{\rm c}_k (\hat b_k+ \hat b^\dag_k)\,.
\ee
Note that, $\sum_{l>m} \omega_{lm} x'_{lm} {\hat P'}_{lm}= ( \sum_{l<m} \omega_{ml} x'^*_{ml} {\hat P'}_{ml})^{\dagger} $.

Finally, applying the RWA, \eqref{VD3} becomes
\be\label{VD4}
\hat H'_{\rm cb} = -i \sum_{m >l}\omega_{lm} { x'}_{lm}  {\hat P'}_{lm}  \sum_k   \eta^{\rm c}_k  b^\dag_k +{\rm H.c.}\, .
\ee

\subsection{Modeling the qubit-bath interaction in the Coulomb and dipole gauge}

Due to the fact that atoms interact with the electromagnetic field, here  we model (as usual) the qubit-bath interaction considering  the environment field as a free-space electromagnetic field described by a collection of harmonic oscillators. We indicate with $\hat c_k$ and $\hat c^\dag_k$ the bosonic photon destruction and creation operators for the $k$-th mode of the reservoir, so that the free reservoir Hamiltonian can be expressed as $\hat H_{c}=\sum_k\omega_k\hat c_k^{\dagger} \hat c_k$.

\subsubsection{Qubit-bath interaction in the Coulomb gauge}\label{app:qubit_bath_interacton_Coulomb}
 
In the  Coulomb gauge, in contrast to the field momenta,  the matter momenta are modified by the light-matter interaction.
In particular, the canonical momenta of the matter field differ from the kinetic momenta (see, e.g., Ref.~\cite{Settineri2021}).
Hence, this gauge is not the more convenient one to study the qubit properties. 

The qubit-bath interaction Hamiltonian in the Coulomb gauge can be obtained by applying the generalized minimal coupling replacement including, in addition to the qubit interaction with the cavity field, also the interaction with the environment. Specifically, the interaction of the qubit with the cavity and the bath fields can be directly obtained by applying to the qubit bare Hamiltonian $\hat H_q = \omega_q \hat \sigma_z /2$ a unitary transformation defined by the following unitary operator
\be\label{App:Gauge_coulomb_trasformation_qubit_bath}
\hat {\cal U}_{qb} = \exp \left\{i \hat \sigma_x [\eta (\hat a + \hat a^\dag) +\sum_k
\eta^{\rm q}_{k} (\hat c_k + \hat c_k^\dag)] 
\right\}\, .
\ee
Keeping only the linear  terms in the qubit-bath coupling strength $\eta_k^q$,
the resulting qubit-bath interaction Hamiltonian in the Coulomb gauge becomes (see Ref.~\cite{Settineri2021})
\be\label{App:Qubit_bath_interaction_coulomb}
\hat {H}_{qb} = \omega_q \hat \Sigma_y \sum_k \eta^{\rm q}_k
(\hat c_k + \hat c_k^\dag) \, ,
\ee
where
\be
\hat \Sigma_y \equiv \hat {\cal R}_{qb}^\dag \hat \sigma_y \hat {\cal R}_{qb} =
\hat \sigma_y {\rm cos}\left[2\eta  (\hat a + \hat a^\dag)\right]-\hat \sigma_z {\rm sin}\left[2\eta  (\hat a + \hat a^\dag)\right]  \, 
\ee
is the Pauli $y$-operator in the dipole gauge, transformed in the Coulomb gauge, and  $\hat {\cal R}_{qb} = \hat {\cal U}_{qb}^\dag$. Analogously, we have 
\begin{align}
\hat \Sigma_z 
&\equiv \hat {\cal R}_{qb}^\dag \hat \sigma_z \hat {\cal R}_{qb} \nonumber \\
&= \hat\sigma_z {\rm cos}\left[2\eta (\hat a + \hat a^\dag)\right]+ \hat \sigma_y {\rm sin}\left[2\eta  (\hat a + \hat a^\dag)\right],  \\
\hat \Sigma_x &\equiv  \hat {\cal R}_{qb}^\dag \hat \sigma_x \hat {\cal R}_{qb}=\hat \sigma_x\, .
\end{align}
A comparison between \eqref{App:Hamiltionian_SB} and \eqref{App:Qubit_bath_interaction_coulomb} clearly shows that, $\hat S=  \omega_q \hat \Sigma_y$ and $\hat B_k= \hat c_k + \hat c_k^\dag$.

By  expanding  the  qubit  operator $\hat \Sigma_y$ in \eqref{App:Qubit_bath_interaction_coulomb}  in  the dressed representation, we obtain
\be\label{Vqd1}
\hat H_{\rm qb} = \omega_q \sum_{m,l} (\Sigma_y)_{lm} \hat P_{lm}
\sum_k \eta^{\rm q}_k (\hat c_k + \hat c_k^\dag) \, ,
\ee
where $(\Sigma_y)_{lm} = \langle l| \hat \Sigma_y |m \rangle$ are the matrix elements of the Pauli $y$-operator, being $|l \rangle$ the energy eigenstates of the Rabi Hamiltonian in the Coulomb gauge [\eqref{eq: coulomb_gauge_rabi_hamiltonian} in \secref{sec: Theoretical_Model_Coulomb_gauge}].
The matrix elements in \eqref{Vqd1} satisfy the relation 
\be
\omega_q (\Sigma_y)_{lm} = i \omega_{ml} (\Sigma_x)_{lm}\,.
\ee

As done for the previously cavity-bath interaction Hamiltonian, we can expand the qubit dressed-operators in terms of positive and negative frequencies operators, such that, applying the RWA, \eqref{Vqd1} can be written as
\be\label{Vqd2}
\hat H_{\rm qb} = i \sum_{m>l} \omega_{ml} (\Sigma_x)_{lm} \hat P_{lm}
\sum_k \eta^{\rm q}_k  \hat c_k^\dag + {\rm H.c.} \, .
\ee

\subsubsection{Qubit-bath interaction in the dipole gauge}

In the dipole gauge, the qubit-bath interaction Hamiltonian can be obtained by applying the generalized minimal coupling replacement to the bosonic Hamiltonian for the bare field-bath $\hat H_b=\sum_k\omega_k\hat c_k^{\dagger} \hat c_k$. 
This corresponds to transforming the bath operators as follows
\be
c_k \to \hat {\cal R}_{qb} \hat c_k \hat {\cal R}^\dag_{qb} = \hat c_k + i\eta^{\rm q}_k \hat \sigma_x\, ,
\ee
where $\hat {\cal R}_{qb} = \hat {\cal U}^\dag_{qb}$.
The resulting qubit-bath Hamiltonian in the dipole gauge becomes
\be\label{App:Qubit_bath_interaction_Dipolo}
\hat H'_{\rm qb} = i \sigma_x \sum_k \omega_k \eta^{\rm q}_k (\hat c_k^\dag - \hat c_k)\, .
\ee
A comparison between \eqref{App:Hamiltionian_SB} and \eqref{App:Qubit_bath_interaction_Dipolo} clearly shows that, $\hat S=  -i \hat \Sigma_x $ and $\hat B_k= \omega_k(\hat c_k - \hat c_k^\dag)$.

By  expanding  $\hat \sigma_x$ in  the dressed representation, we obtain
\be\label{Vqbd1}
\hat H'_{\rm qb} = -i \sum_{m,l}(\sigma_x)'_{lm} \hat P'_{lm}\sum_k \omega_k \eta^{\rm q}_k (\hat c_k - \hat c_k^\dag)\, ,
\ee
where $(\sigma_x)'_{lm} = \langle l'| \hat \sigma_x |m' \rangle$ are the matrix elements of the Pauli $x$-operator, being $|l' \rangle$ the energy eigenstates of the Rabi Hamiltonian in dipole gauge [\eqref{eq: dipole_gauge_rabi_hamiltonian} in \secref{sec: Theoretical_Model_dipole_gauge}].

Finally, in terms of positive and negative frequencies of the qubit dressed-operators and applying the RWA, \eqref{Vqbd1} can be written as
\be
\hat H'_{\rm qb} = i \sum_{m'>l'}(\sigma_x)'_{lm} \hat P'_{lm}\sum_k \omega_k \eta^{\rm q}_k  \hat c_k^\dag + {\rm H.c.}\, .
\ee
\vspace{0.2cm}

\section{Gauge invariance of the generalized master equation}
\label{app: generalized master equations}

The GME \cite{Settineri2018} for strongly interacting hybrid quantum systems is developed using the basis of the system eigenstates (dressed-states). Moreover, it does not makes use of the secular approximation (post-trace rotating wave approximation). Thus, unlike the dressed master equation in Ref.~\cite{Beaudoin2011}, suitable for anharmonic systems, it is also valid for systems displaying harmonic or quasi-harmonic spectra. This feature is essential to describe the system dynamics and the resulting spectra both at small and at very large (deep strong) values of the normalized coupling strength $\eta$. Here, we provide a demonstration of the GME gauge-invariance. Within this approach, we guarantee the gauge invariance of matrix elements of the density matrix and expectation values, as well as the obtained emission rate and spectra.

We start considering the GME in the Coulomb gauge for the Rabi model, which can be written as
\be
\label{eq: generalized master equation}
\dot{\hat{\rho}} = -i \comm{\hat{H}_R}{\hat{\rho}} + \mathcal{L}_{\text{g}} \hat{\rho}\;,
\ee
where the dissipator $\mathcal{L}_{\text{g}}$ contains two contributions $\mathcal{L}_{\text{g}}=\mathcal{L}^{\rm c}_{\rm g} + \mathcal{L}^{\rm q}_{\rm g}$, arising from the cavity-bath (c) and the qubit-bath (q) interaction, respectively:
\begin{widetext}
\bea
\mathcal{L}^{\rm c(q)}_{\rm g} \hat \rho &=& \frac{1}{2} \sum_{\substack{j > k \\ l > m}} \left\{
\Gamma^{\rm c(q)}_{lm} n_{ml}({\cal T}_{c(q)}) \left[ \hat P_{lm} \hat \rho \hat P_{kj}
-  \hat P_{kj} \hat P_{lm} \hat \rho   \right] +
\Gamma^{\rm c(q)}_{kj} n_{jk}({\cal T}_{c(q)}) \left[ \hat P_{lm} \hat \rho \hat P_{kj}
-  \hat \rho \hat P_{kj} \hat P_{lm}  \right]
\right. \nonumber\\
&+& \left. \Gamma^{\rm c(q)}_{lm}[n_{ml}({\cal T}_{c(q)}) +1] \left[ \hat P_{kj} \hat \rho \hat P_{lm}
-  \hat \rho \hat P_{lm} \hat P_{kj}  \right] + 
\Gamma^{\rm c(q)}_{kj}[n_{jk}({\cal T}_{c(q)}) +1]
\left[ \hat P_{kj} \hat \rho \hat P_{lm}
-  \hat P_{lm} \hat P_{kj} \hat \rho   \right] \right\}\, ,
\label{GME}\eea
\end{widetext}
with
\be
\label{NT}
n_{kj}({\cal T}_{c(q)}) = \{ \exp [\omega_{kj} / (\omega_{c(q)}{\cal T}_{c(q)})] - 1\}^{-1}\, ,
\ee
which is 
the thermal populations of the cavity and qubit reservoirs calculated at the transition frequencies $\omega_{kj}$.

The cavity- and qubit-bath coupling rates determine both the losses and the incoherent pumping of the system. These rates depend on the specific interaction Hamiltonian (which describe the coupling between the system component and its bath) and on the density of the states of the relative bath. 

For the cavity-bath rate, following the general derivation in \appref{app:cavity_bath_interaction_coulomb}, and using \eqref{VC2n}, we obtain
\be\label{Gc}
\Gamma^{\rm c}_{kj} = 2 \pi d_c(\omega_{kj}) |\omega_{kj} \eta^{\rm c} (\omega_{kj})|^2 |\langle k|(\hat a + \hat a^\dag) | j \rangle|^2\, ,
\ee 
where $d_c(\omega_{kj})$ is the density of states of the bath calculated at the transition frequency $\omega_{kj}$.
For any numerical calculations in the main text, we assumed $\eta^{\rm c}_k \propto 1/\sqrt{\omega_k}$, and  $d_c$ independent on $\omega$. Taking these assumptions into account, and including all the constants in the rate $\kappa$, the cavity-bath coupling rate can be expressed  as
\be\label{Gc1}
\Gamma^{\rm c}_{kj} = \kappa \frac{\omega_{kj}}{\omega_c}\left |\langle j|( \hat a + \hat a^\dag)  |k \rangle \right|^2\,  .
\ee
It is clear that the cavity-bath interaction rates in \eqref{Gc} and \eqref{Gc1} are gauge invariant.
Analogously, from the general derivation in \appref{app:qubit_bath_interacton_Coulomb}, and using \eqref{Vqd2} we obtain
\be\label{Gq}
\Gamma^{\rm q}_{kj} = 2 \pi d_q(\omega_{kj}) |\omega_{kj} \eta^{\rm q} (\omega_{kj})|^2 |\langle k| \hat \sigma_x | j \rangle|^2\, .
\ee
Assuming $\eta^{\rm q}_k \propto 1/\sqrt{\omega_k}$, $d_c$ to be independent on $\omega$, and including all the constants in the rate $\gamma$, the qubit-bath coupling rate can be expressed  as
\be\label{Gq1}
\Gamma^{\rm q}_{kj} = \gamma \frac{\omega_{kj}}{\omega_q}\left|\langle j| \hat \sigma_x |k \rangle \right|^2\,  .
\ee

\subsection{Gauge invariance}\label{app: generalized master equations_demostration}

In any coherent description of the light-matter interaction, the expectation value of any Hermitian operator $\hat O$ that characterizes a property of the optical system, has to be gauge invariant. Since  the correct gauge transformation is described by a unitary operator $\hat {\cal U}$, if $|\psi \rangle$ is a quantum state in the Coulomb gauge, it is trivial to see that
\be\label{App:invariant}
\langle \psi | \hat O | \psi \rangle = \langle \psi' | \hat O' | \psi' \rangle\, ,
\ee
where $\hat O' = \hat {\cal U} \hat O \hat {\cal U}^\dag$ and $| \psi' \rangle = \hat {\cal U} |\psi \rangle$
are a generic Hermitian operator and a ket state in the dipole gauge, respectively. In particular, by considering the definition of the density operator $\hat \rho = \sum_m p_m |\psi_m \rangle \langle \psi_m|$ one can clearly see that, while the matrix elements $p_m$ are gauge-invariant probabilities the quantum states $|\psi_m \rangle$ are gauge-dependent.  If $|\psi_m \rangle$ are obtained in the Coulomb gauge, in the dipole gauge we have
\be
\hat \rho' = \sum_m p_m |\psi'_m \rangle \langle \psi'_m| = \hat {\cal U} \hat \rho\,
\hat {\cal U}^\dag 
\ee

Here, we show that the master equation in \eqref{GME} does not break gauge-invariance. As done so far, we are labeling quantum states and operators in the dipole gauge with a prime apex primed superscript.

Due to the \eqref{App:invariant}, we want to show that the time-evolution  of any system operator is gauge-invariant, namely
\be\label{GI}
\frac{d}{dt} \langle \hat O \rangle = \Tr{\hat O\,  \dot {\hat \rho}} = 
\Tr{\hat O' \dot {\hat \rho}' }'.
\ee
In terms of \eqref{eq: generalized master equation}, we can write \eqref{GI} in the Coulomb gauge as
\bea\label{Gaugero}
\Tr{\hat O \dot{\hat \rho}  } &=& 
-i \Tr{\hat O [\hat H_R, \hat \rho]} 
 + \Tr{\hat O {\cal L}_{\rm g} \hat \rho}
 \\
&=&-i \sum_j \langle j | \hat O [\hat H_R, \hat \rho]  |j\rangle + \sum_j \langle j | \hat O  {\cal L}_{\rm g} \hat \rho  |j\rangle\, . \nonumber
\eea

Let us next consider separately the two terms on the right hand side of \eqref{Gaugero}. By means of the unitary gauge transformation $\hat {\cal U}$
, we obtain
\begin{widetext}
\bea
\sum_j \langle j | \hat O ( \hat H_R \hat \rho - \hat \rho \hat H_R ) |j\rangle
&=& \sum_j \langle j' |[ \hat {\cal U}  \hat {\cal U}^\dag (\hat {\cal U}\hat O \hat {\cal U}^\dag) [ (\hat {\cal U} \hat H_R \hat {\cal U}^\dag) (\hat {\cal U} \hat \rho \hat {\cal U}^\dag)- (\hat {\cal U} \hat \rho \hat {\cal U}^\dag) (\hat {\cal U}\hat H_R \hat {\cal U}^\dag)\hat {\cal U} \hat {\cal U}^\dag]  |j'\rangle \nonumber \\
&=& \sum_j \langle j' | \hat O' ( \hat H'_R \hat \rho' - \hat \rho' \hat H'_R)  |j'\rangle = \Tr{\hat O' [\hat H'_R, \hat \rho']}'\, .
\eea
\end{widetext}
To deal with the second term on the right hand side of \eqref{Gaugero}, we first observe that $\Gamma^{\rm c(q)}_{ml} = \Gamma^{\rm c(q)}_{m'l'}$ and $n_{ml}({\cal T}_{c(q)}) = n_{m'l'}({\cal T}_{c(q)})$ in \eqref{GME} are gauge invariant. 
Thus, it is sufficient to show that the terms in the square brackets (involving the transition operators $\hat P_{lm}$) in \eqref{GME}, provide gauge invariant contributions in  \eqref{GI}. Considering the first of  these terms, we have
\begin{widetext}
\bea
\Tr{ \hat O \hat P_{lm} \hat \rho \hat P_{kj}} &=& \sum_n \langle j| 
\hat O \hat P_{lm} \hat \rho \hat P_{kj}
| j \rangle =  \sum_n \langle j'| \hat {\cal U} \hat {\cal U}^\dag
(\hat {\cal U}\hat O \hat {\cal U}^\dag)
(\hat {\cal U} \hat P_{lm} \hat {\cal U}^\dag)
(\hat {\cal U}\hat\rho \hat {\cal U}^\dag)
(\hat {\cal U} \hat P_{kj} \hat {\cal U}^\dag)
\hat {\cal U} \hat {\cal U}^\dag
| j' \rangle  \\
&=& \sum_n \langle j'| \hat O' \hat P_{lm} \hat \rho' \hat P_{kj}
| j' \rangle
= \Tr{ \hat O' \hat P'_{lm} \hat \rho' \hat P'_{kj}}'\, . \nonumber
\eea
\end{widetext}
It is easy to see that the other similar terms transform exactly in the same way.
This proves that
\[\Tr{\hat O {\cal L}_{\rm g} \hat \rho} = \Tr{\hat O' {\cal L}'_{\rm g} \hat \rho'}'\, ,
\]
and, as a consequence,  \eqref{GI} is  gauge invariant.

\clearpage

\bibliography{Riken}

\end{document}